\documentclass[reprint,superscriptaddres,aps,bbm,pre,longbibliography,floatfix]{revtex4-1}
\usepackage{graphics}
\usepackage{bm}
\usepackage{lipsum}
\usepackage{epstopdf}
\usepackage{graphicx}
\usepackage{amsmath}
\usepackage{amsthm}
\usepackage{bm}
\usepackage{bbm}
\usepackage{amssymb}
\usepackage{physics}
\usepackage{braket}
\usepackage{lipsum}
\usepackage{comment}
\begin{document}
\title{History state formalism for time series with application to finance} 
\author{F.\ Lomoc$^1$,  N. Canosa$^1$, A.P.\ Boette$^1$, R. Rossignoli$^{1,2}$}
\affiliation{$^1$ Instituto de F\'{\i}sica La Plata, CONICET, and Departamento de F\'{\i}sica, Universidad Nacional de La Plata,
C.C. 67, La Plata (1900), Argentina\\
$^{2}$ Comisi\'on de Investigaciones Cient\'{\i}ficas (CIC), La Plata (1900), Argentina}

\begin{abstract}
We present a method for analyzing general time series by employing the history state formalism of quantum mechanics. This formalism allows us to describe a  complete evolution 
based on a single quantum state, the history state, which simultaneously includes -also as a quantum system- the reference clock. It naturally leads to the concept of system-time entanglement, with the ensuing entanglement entropy constituting  a measure of the effective number of distinguishable states visited in the history. Through a quantum coherent state embedding of the time series data, it is then possible to associate a quantum history state to the series.  The gaussian overlap between these coherent states provides thus a smooth measure of distinguishability between the series data. The eigenvalues of the corresponding overlap matrix determine in fact the entanglement spectrum and  entropy of the history state, 
which provide a rigorous characterization of the evolution.   
As illustration, the formalism is applied to   typical financial time-series data. 
Through the  entanglement entropy and spectrum, 
different evolution regimes can be identified.  Entanglement based volatility indicators are also derived, and  compared with standard  volatility measures.
\end{abstract} 

\maketitle

\section{Introduction}

Recent investigations in the area of Quantum Mechanics (QM) 
and Quantum Information 
have led  to  the development of the history state formalism \cite{Paw2.83,
Ma.15, Moreva.14, Clean.15,BRGC.16,BR.18}, which incorporates time in an entirely quantum embedding. 
Starting from a relational state originally proposed by Page and Wootters  \cite{Paw2.83}, the formalism includes time through a reference clock, 
which is another quantum system itself, whose states serve as a time reference for the first system. Then the whole system can be  described through a system-time history state, a joint quantum state of both the system and the  clock, whose entanglement (``system-time entanglement'') indicates the ``degree'' of  evolution of the system in time \cite{BRGC.16,BR.18}. Moreover, the system evolution equation  itself, i.e., the Schr\"odinger equation, can also be derived from the joint history state,  when forced to  satisfy a proper ``static'' eigenvalue equation \cite{Paw2.83,Ma.15,BR.18}. 
The  formalism  has recently gained  attention in  nonrelativistic  QM \cite{Ma.15,Moreva.14,Clean.15,BRGC.16,Pa.17,BR.18,Ni.18,Ll.19,DP.19,Lo.22,DN.25},  and has been also extended  to the relativistic regime  \cite{zh.18,di.19,Diaz.212,GLM.23,ND.24}.
 
The aim of this  work is to further this path, by studying time series from the perspective of  the  history state framework, where fully discrete formulations   have also been considered \cite{Clean.15,BRGC.16,BR.18,Lo.22,DN.25}. We show that by  associating the time series   to a series of coherent quantum states, it is possible to define a  particular history state, which describes important properties of the complete time series evolution. Its entanglement entropy  is a measure of the number of effective orthogonal states visited during  the evolution \cite{BRGC.16}, then quantifying the degree of  distinguishability of the set of values taken in the series. The more distinct the states are, i.e.\ the more variability is detected in the evolution, the more closely to orthogonal will the different states be, which translates into more entanglement between the system and the clock. The associated entanglement entropy depends in fact just on the 
  overlaps between the states visited in the whole history, i.e., on the eigenvalues of the ensuing overlap matrix, which determine the entanglement spectrum.      
  As  will be shown, 
  this  spectrum provides  further insights  for characterizing the evolution of the time series, supported by  the rigorous  perspective  of majorization theory \cite{Bh.97,MOA.11}.

A fundamental aspect is that through the present coherent state quantum embedding of time series data (and a conveniently chosen scale factor), their basic gaussian nonorthogonality  allows one to represent continuously the degree of distinguishability of data values, such that  
closely lying data whose difference is just barely relevant or  slightly resolved correspond to states with high overlap. Hence, rather than a sharp cutoff, quantum states enable a consistent smooth transition between coincident and clearly distinguishable data. Moreover, through the  Schmidt  decomposition of the history state and the associated entanglement spectrum, the formalism 
provides  a rigorous   picture of the evolution in terms of a set of effective orthogonal states visited in the history and a concomitant  set of positive  probabilities or ``occurrence frequencies'', 
which are just these eigenvalues. 
We note here that an arbitrary ``overlap function'' not based on true quantum states (i.e., on a true Gram or overlap matrix) does not ensure positive eigenvalues. On the other hand, a quantum embedding also allows in principle to quantum-simulate a time series, which opens the way to possible quantum computation \cite{Nielsen.01} based schemes. 

As specific application, we will  consider the 
analysis through this formalism  of financial time series. In recent decades, there has been a growing body of applications of mathematical and physical methods in the area of finance, leading to the establishment of what is called quantitative finance \cite{black,merton,elliott2005mathematics,shreve2005stochastic,WHY.06,FPW.11,muldowney2013modern,Paw.14,BAM.17}. In addition,  the interdisciplinary field called ``Quantum Finance'', which merges methods and formalisms stemming from QM and quantitative finance, has provided a new perspective for the treatment of diverse problems in the area that has been receiving increasing  attention \cite{schaden2002quantum,haven02,chen2001quantum,baaquie07, Qmodel.2010,contreras2010quantum,PRA.2018,orus.19,orrell.2020,egger20,lu.2024,qwfinan25}. 

Here we will examine well known financial time series.    
We start with the SPY, which 
is the ticker symbol for an ETF (Exchange-Traded Funds) 
that tracks the performance of the S\&P500 (Standard and Poor's) index. The time series  of the gold ETF (GLD) and Bitcoin (BTC) will be also considered. By means 
of the system-time entanglement entropy and spectrum, we will  characterize different regimes of prices along the evolution. Besides, we will derive  measures that are able to recognize the variability of the asset prices and that can be interpreted as  proxies of their volatility. They will be compared  with  standard  volatility measures for these assets, namely the volatility index (VIX) \cite{VIX.09}  for the SPY and GLD, and the Crypto Volatility Index (CVI) \cite{CVI} for Bitcoin.

We first  review in sections \ref{IIA}--\ref{IIC} the  history state formalism, including the concept of system-time entanglement entropy and its relation with mutual information, while  generalized entropies, their relation with majorization theory and their application to the 
characterization of different evolution regimes of entanglement entropies,  are  described in \ref{IID}.  In \ref{IIE}--\ref{IIF} we introduce  the history state formalism for time series through the quantum coherent state embedding of the time series data, 
and the ensuing gaussian overlap. 
 Finally, in section \ref{III}  we illustrate the  formalism  through its application to financial time series for the above mentioned assets, where different types of history states are analyzed. Conclusions are drawn in \ref{IV}.  
  App.\ \ref{ApA} contains the essential  elements of majorization theory and related demonstrations of results in \ref{IID}, while  App.\  \ref{ApB} provides a summary of the main features of volatility measures. 
 
\section{Formalism} \label{II}

\subsection{Discrete history state formalism in QM \label{IIA}}
 We consider a  quantum system  $S$ whose states $|\psi\rangle$ belong to a Hilbert space 
 $\mathcal{H}_S$, 
  and  a  reference  clock system $T$ 
 with Hilbert space $\mathcal{H}_T$. If $|\psi_n\rangle\in {\cal H}_S$ is the state  of $S$ when  the clock is in state $|n\rangle\in{\cal H}_T$, the history state  $|\Psi\rangle \in {\cal H}_S\otimes {\cal H}_T$ over $N$ distinguishable `times'  $|n\rangle$, $n=0,\ldots,N\!-\!1$,  $\langle n|n'\rangle=\delta_{nn'}$, is 
 \begin{equation}|{\Psi}\rangle =\tfrac{1}{\sqrt{N}}\sum_{n=0}^{N-1} |\psi_n\rangle\otimes |n\rangle\,.\label{t0}
\end{equation}
It describes the whole evolution of  $S$ along the chosen set of $N$ times. The states  $|\psi_n\rangle$ are normalized but obviously not necessarily orthogonal, and can be recovered from $|\Psi\rangle$ as 
$|\psi_n\rangle=\sqrt{N}\langle n|\Psi\rangle$, where 
 the ``partial'' product  is defined through $\langle n|(|\psi_S\rangle\otimes|\phi_T\rangle)=\langle n|\phi_T\rangle |\psi_S\rangle$. 
 
  All information about the system evolution can then be retrieved from $|\Psi\rangle$.  In particular, time averages of system  $S$ observables $O_S$,   defined as 
 \begin{equation}
 \langle O_S\rangle_N=\tfrac{1}{N}\sum_{n=0}^{N-1} \langle\psi_n|O_S|\psi_n\rangle\,,\label{OST}
 \end{equation}
 which in principle require averages (and hence measurements) at each time $n$, can be directly obtained as 
 \begin{equation}
 \langle O_S\rangle_N=\langle\Psi|{\cal O}_S|\Psi\rangle\,,\label{OST2}
  \end{equation}
where ${\cal O}_S=O_S\otimes \mathbbm{1}_T$,  
such that $\langle n| {\cal O}_S|n'\rangle=O_S\delta_{nn'}$. Thus, a realization  of $|\Psi\rangle$   
enables an efficient evaluation of time averages and related quantities as a single expectation value \eqref{OST2} \cite{BRGC.16,BR.18,DN.25}.   

We can always consider the normalized states $|\psi_n\rangle$ as emerging from an initial state $|\psi_0\rangle$ through a unitary evolution,  \begin{equation}|\psi_n\rangle=U_n|\psi_0\rangle\,,\end{equation}
with $U_n^\dag U_n=\mathbbm{1}_S$ 
and $U_0=\mathbbm 1_S$. Then we can define a cyclic system-clock  translation superoperator 
\begin{equation}
{\cal U}=\sum_{n=1}^{N} U_{n,n-1}\otimes |n\rangle\langle n-1|,
\label{u}
\end{equation}
over ${\cal H}_S\otimes {\cal H}_T$, where $U_{n,n-1}=U_{n}U_{n-1}^\dag$ and     $U_{N}\equiv U_{0}$, $|N\rangle\equiv |0\rangle$,  such that ${\cal U}^\dag {\cal U}=\mathbbm 1_S\otimes \mathbbm 1_N$ ($\mathbbm 1_N=\sum_{n=0}^{N-1}|n\rangle\langle n|$ is the  projector onto the subspace of ${\cal H}_T$ spanned by the $N$ clock states $|n\rangle$) and  ${\cal U}\,|\psi_{n-1}\rangle\otimes|n\!-\!1\rangle$ $=|\psi_n\rangle\otimes |n\rangle$ $\forall\,n$. Hence,  the history state \eqref{t0} satisfies 
\begin{equation}
{\cal U}|\Psi\rangle=|\Psi\rangle\,,\label{t2}
\end{equation}
 i.e.\ it is an exact eigenstate of ${\cal U}$ with eigenvalue $1$, thus remaining  fully  
invariant under ${\cal U}$. 

Conversely,  Eq.\ \eqref{t2}, which can be considered as a discrete version of a Wheeler-DeWitt-type equation \cite{dewitt.1967}, {\it implies} that $|\Psi\rangle$ is necessarily a history state of the form \eqref{t0}, with $|\psi_n\rangle=U_n|\psi_0\rangle$ undergoing a unitary stepwise evolution and  $|\psi_0\rangle$ an {\it arbitrary} normalized initial state: Eqs.\ \eqref{u}--\eqref{t2} entail that just the $N$ times $|n\rangle$ enter in  the support of $|\Psi\rangle$, such that $|\Psi\rangle=\sum_n \langle n|\Psi\rangle\otimes |n\rangle$,  then having the  form $\eqref{t0}$ for $|\psi_n\rangle=\sqrt{N}\langle n|\Psi\rangle$, with  
 $\langle n|\Psi\rangle=\langle n|{\cal U}|\Psi\rangle=U_{n,n-1}\langle n\!-\!1|\Psi\rangle$, i.e., $|\psi_{n}\rangle=U_{n,n-1}|\psi_{n-1}\rangle=U_n|\psi_0\rangle$,   
and  $\langle \Psi|\Psi\rangle=\frac{1}{N}\sum_n\langle \psi_n|\psi_n\rangle=\langle \psi_0|\psi_0\rangle=1$  if $|\Psi\rangle$ is assumed normalized.  
 Thus, unitary stepwise evolution of system $S$ states 
$\propto \langle n|\Psi\rangle$ directly follows from Eqs.\ \eqref{u}--\eqref{t2} \cite{BRGC.16, BR.18}. 

The history state can itself be generated from an initial seed  state $|\psi_0\rangle\otimes |0\rangle$ 
through a unitary quantum gate ${\cal W}$, 
\begin{equation}
    |\Psi\rangle={\cal W}(|\psi_0\rangle\otimes |0\rangle)\,,\label{gen}
\end{equation}
    where ${\cal W}=\tilde {\cal U}({\mathbbm 1}_S\otimes H_N)$. Here $H_N$ is a  Hadamard-like local operator on ${\cal H}_T$ satisfying $H_N|0\rangle=\frac{1}{\sqrt{N}}\sum_n |n\rangle$,  whereas  $\tilde {\cal U}=\sum_n  U_n\otimes \ketbra{n}{n}$ is a control-like operator on ${\cal H}_S\otimes {\cal H}_T$, such that 
    ${\cal U}=\tilde{\cal  U}(\mathbbm 1\otimes e^{-iP})\tilde{\cal U}^\dag$, with $e^{-iP}=\sum_n\ketbra{n}{n\!-\!1}$ a translation operator in  ${\cal H}_T$.
  A history over $N=2^{m}$ times can then be efficiently achieved {\it with just $m$ qubits}, in which case $H_N=H^{\otimes m}$, with $H$ the single-qubit Hadamard operator (see refs.\ \cite{BRGC.16,BR.18,DN.25}).
 
 \subsection{System-time entanglement as quantifier of evolution\label{IIB}}
The history state \eqref{t0} is in general an {\it entangled} state, i.e., it is not a product state $|\psi\rangle_S\otimes |\phi\rangle_T$.  The only exception is the  case of {\it stationary} system states,  i.e.\ states which evolve just with a phase,   
$|\psi_n\rangle 
=e^{-i \phi_n}|\psi_0\rangle$ $\forall\,n$  for which it becomes  {\it separable}: $|\Psi\rangle=|\psi_0\rangle\otimes|\phi\rangle_T$, with $|\phi\rangle_T=\frac{1}{N}\sum_n e^{-i\phi_n}|n\rangle$ a superposition of all time states $|n\rangle$.  In this case there is no entanglement between  $S$ and the clock $T$, i.e., between ``system and time''. 

The opposite limit is that in which the system   evolves into a new orthogonal state at each step, such that $\langle\psi_n|\psi_{n'}\rangle=\delta_{nn'}$ $\forall\,n,n'$. In this case $|\Psi\rangle$  becomes {\it maximally entangled}, with \eqref{t0} becoming its Schmidt decomposition. Therefore, its ``degree of entanglement'', i.e.\ of the system-time entanglement, is a measure of the distinguishable evolution of the system $S$ along the $N$ times. 

We now recall that bipartite pure state  entanglement can be measured through the entanglement entropy,  
\begin{equation} E(S,T)=S(\rho_S)=S(\rho_T)
\label{EST}\,,
\end{equation}
where $\rho_{S(T)}$  are the  reduced states of  $S$ ($T$),  
\begin{subequations}
\begin{eqnarray}
\rho_S&=&{\rm Tr}_T\,|\Psi\rangle\langle\Psi|=\frac{1}{N}\sum_n|\psi_n\rangle\langle\psi_n|\,,\label{rhs}\\
\rho_T&=&{\rm Tr}_S\,|\Psi\rangle\langle\Psi|=\frac{1}{N}\sum_{n,n'}\langle\psi_{n'}|\psi_n\rangle |n\rangle\langle n'|\,,\label{rht}
\end{eqnarray}
\label{rhST}
\end{subequations}
$\!\!$with ${\rm Tr}_{T(S)}$ denoting  the partial trace. These reduced states  are  {\it isospectral}, i.e., they have the same nonzero eigenvalues and hence the same entropy. This set of eigenvalues is the {\it entanglement spectrum}.  The entropy in \eqref{EST} will be first chosen as the  von Neumann entropy 
  \begin{equation}S(\rho)=-{\rm Tr}\,\rho\log\rho\,.\label{S1}\end{equation} 
  
 Explicitly, we see from Eq.\ \eqref{rht} that 
the eigenvalues of $\rho_T$ 
are just those of the {\it  normalized  overlap matrix} 
${\bf O}_N$ of the system states $|\psi_n\rangle$, of elements
\begin{equation}({\bf O}_N)_{n'n}=\langle n|\rho_T|n'\rangle=\langle\psi_{n'}|\psi_n\rangle/N\,,\label{O}\end{equation}
which is 
hermitian and positive semidefinite, satisfying ${\rm Tr}\,{\bf O}_N=1$. Since it is the Gram matrix of the states $|\psi_n\rangle/\sqrt{N}$, its rank $N_s$, i.e., the number of nonzero eigenvalues, is just   {\it the number of  linearly independent  system states $|\psi_n\rangle$  involved in the discrete evolution,    i.e., the dimension of the subspace ${\cal S}\subset{\cal H}_S$ spanned  by the $N$ states $|\psi_n\rangle$}.  
If ${\bf W}$ is the unitary  matrix diagonalizing ${\bf O}_N$,  \begin{equation}({\bf W}^\dag {\bf O}_N {\bf W})_{kk'}=\lambda_k\delta_{kk'}\,,\end{equation} 
with $\lambda_k\geq 0$, $\sum_k \lambda_k=1$, we may rewrite the history state \eqref{t0} and the  densities  \eqref{rhs}--\eqref{rht} in the diagonal forms 
\begin{eqnarray}
|\Psi\rangle&=&\sum_{k=1}^{N_s} \sqrt{\lambda_k}\,|k\rangle_S\otimes |k\rangle_T\,,\label{SD}\\
\rho_S&=&\sum_{k=1}^{N_s}\lambda_k|k\rangle_S\langle k
|\,,\label{rhstd}\;\;
\rho_T=\sum_{k=1}^{N_s} \lambda_k |k\rangle_T\langle k|\,,\end{eqnarray}
where sums run just over the $N_s$ non-zero eigenvalues $\lambda_k>0$ of ${\bf O}_N$  and \begin{equation}|k\rangle_S=\frac{1}{\sqrt{N\lambda_k}}\sum_n W_{nk}|\psi_n\rangle\,,\;\; 
|k\rangle_T=\sum_n W_{nk}^*|n\rangle\,,\label{ks}\end{equation} 
are {\it orthogonal} states of $S$ and $T$: $_{S}\langle k'|k\rangle_{S}=\delta_{kk'}$ for $k\leq N_s$, 
$_T\langle k|k'\rangle_T=\delta_{kk'}$ for $k\leq N$. 
Eq.\ \eqref{SD} is precisely the {\it Schmidt decomposition} of $|\Psi\rangle$. 

The  entropy \eqref{EST} can then be explicitly written as 
\begin{equation}
E(S,T)=-\sum_{k=1}^{N_s}\lambda_k \log \lambda_k\,,\label{EST2}
\end{equation}
satisfying $0\leq E(S,T)\leq \log N$. 
 The lower limit is reached in the stationary case $N_s=1$, where $\rho_S=|\psi_0\rangle\langle\psi_0|$, $\rho_T=|\phi\rangle_T\langle\phi|$ are pure and hence $S(\rho_S)=S(\rho_T)=0$,    
 and the upper limit in the maximally evolving case $N_s=N$ and 
$\lambda_k=1/N$ $\forall\,k$,  where $\rho_S=\Pi_S/N$, $\rho_T=\mathbbm 1_N/N$ are maximally mixed, with $\Pi_S$ the  orthogonal projector on the subspace spanned by the  $N$ orthogonal states $|\psi_n\rangle$, and hence 
$S(\rho_S)=S(\rho_T)=\log N$.

Thus, using $\log\equiv\log_2$, we may  approximately consider 
\begin{equation}
N_E=2^{E(S,T)},\label{NS}
\end{equation}
as the {\it effective} number of orthogonal states visited in the evolution along the $N$ times. It is  $1$ in the separable case $E(S,T)=0$ and $N$ in the maximally evolving case $E(S,T)=\log_2 N$, 
lying otherwise in the interval $(1,N)$. 

The effective number \eqref{NS} takes  into account the relative frequency with which distinct states appear. Hence, it is lower than the rank $N_s$ of the overlap matrix, which just counts the  number of linearly independent states of the history regardless of their  
 relative frequencies. 

For example, 
if just $M\leq N$ strictly orthogonal states $|\tilde\psi_m\rangle$ are visited in the $N$ times history, each $l_m$ times  with $\sum_{m=1}^M l_m=N$,
the history state \eqref{t0} becomes  
\begin{equation}
|\Psi\rangle=\sum_{m=1}^M \sqrt{p_m}\,|\tilde\psi_m\rangle\otimes|\tilde m\rangle,\label{Psim}
\end{equation}
with $p_m=l_m/N$ ($\sum_m p_m=1$), 
$\langle \tilde\psi_{m'}|\tilde\psi_{m}\rangle=\delta_{mm'}$ and $|\tilde m\rangle=\frac{1}{\sqrt{l_m}}\sum_n|n\rangle\langle \psi_m|\psi_n\rangle$  normalized orthogonal clock states.  In this case Eq.\ \eqref{Psim} is already the Schmidt decomposition \eqref{SD}, i.e. $\lambda_k=p_m$ in Eq.\ \eqref{EST2}. Then  Eq.\ \eqref{NS}  yields $N_E\leq M$, with 
$N_E=M$  iff all $l_m$ are equal, i.e.\ if all $M$ orthogonal  states are visited the same number of times, such that $l_m=N/M$ and $p_m=1/M$ $\forall\,m$, implying $E(S,T)=\log M$. Otherwise, 
$E(S,T)$ and hence $N_E$ are smaller, 
the minimum reached when just one of the $M$ orthogonal states appears more than once ($p_m=\frac{N-M+1}{N}$ for this state,  $p_m=\frac{1}{N}$ for the rest).

So far we have implicitly assumed ${\rm dim}\,{\cal H}_T\geq N$, the formalism valid for {\it any} dimension $d_S$ of ${\cal H}_S$ (finite or infinite). Nonetheless, if $d_S<N$,  
the rank $N_s$ obviously satisfies $N_s\leq d_S<N$, and it is more 
convenient to write $|\psi_n\rangle=\sum_i C_{in}|i\rangle$, with $\{|i\rangle\}$ any orthogonal basis of ${\cal H}_S$ and $C_{in}=\langle i|\psi_n\rangle$. Then the $N\times N$ overlap matrix ${\bf O}=C^\dag C/N$ will have the same nonzero eigenvalues as the smaller $d_S\times d_S$ matrix $CC^\dagger/N$,  which is thus more convenient. The Schmidt decomposition \eqref{SD} can then be obtained from the singular value decomposition of $C$, with $\lambda_k$ the square of its singular values $\sigma_k$. 

\subsection{Entanglement entropy and mutual information\label{IIC}}
The entanglement entropy \eqref{EST2} also
admits other interpretations. Let us first recall that the mutual information of a bipartite system $A+B$ is defined as 
\begin{eqnarray}
I(A,B)&=&S(A)+S(B)-S(A,B)\,,\label{IM}
\\ &=&S(A)-S(A|B),\label{IM2}
\end{eqnarray}
where $S(A|B)=S(A,B)-S(B)$ is the conditional entropy and in the quantum case, $S(A)=S(\rho_A)$, $S(B)=S(\rho_B)$ and $S(A,B)=S(\rho_{AB})$, 
with  $\rho_{A(B)}={\rm Tr}\,_{B(A)}\,\rho_{AB}$ the reduced density operator of $A$ ($B$).  Eq.\ \eqref{IM} is  a measure of the total (classical plus quantum)  correlations in the system, satisfying $I(A,B)\geq 0$, with $I(A,B)=0$ iff $\rho_{AB}=\rho_{A}\otimes\rho_{B}$. 

For a pure quantum state $\rho_{AB}=|\Psi\rangle\langle\Psi|$, $S(\rho_{AB})=0$ while  $S(A)=S(B)=E(A,B)$, 
with $E(A,B)$ the entanglement entropy (i.e., Eq.\ \eqref{EST} in the case of the history state), such that $I(A,B)=2E(A,B)$.  
Thus, in pure entangled states it exceeds the classical upper bound $I(A,B)\leq {\rm Min}[S(A),S(B)]$ valid for any classical joint  random variable (where $S(A|B)=\sum_j S(A|j)p(B\!=\!j)\geq 0$), as well as for any separable quantum state \cite{NK.01}, since pure state entanglement leads to $S(A|B)=-S(B)=-S(A)<0$. 

Let us now suppose that some decoherence process takes place such that the coherence of the pure history state \eqref{t0} is lost and one is left with the decohered {\it classically correlated state} 
\begin{equation}
\rho'_{ST}=\sum_n \Pi_n\rho_{ST}\Pi_n=\tfrac{1}{N} \sum_n |\psi_n\rangle\langle\psi_n|\otimes |n\rangle\langle n|\,,\label{rhopst}\end{equation}
where $\Pi_n=\mathbbm 1_S\otimes |n\rangle\langle n|$ are the orthogonal projectors on the clock states $|n\rangle$. For example, a local projective measurement in the clock basis  $\{|n\rangle\}$ (without postselection) leads to the postmeasurement state \eqref{rhopst}.
This state is a mixture of product states and hence it is {\it separable} (in the sense that it can be generated by LOCC),  having no entanglement: $E'(S,T)=0$. 

Nonetheless,  the previous entanglement entropy \eqref{EST2} reemerges exactly in \eqref{rhopst} as  {\it mutual information}, now measuring the classical correlation between $S$ and $T$:  In the state \eqref{rhopst} we obtain 
\begin{eqnarray}
I'(S,T)&=&S(\rho'_S)+S(\rho'_T)-S(\rho'_{ST})
\label{Ipst}\\
&=&S(\rho_S)=E(S,T)\,,
\end{eqnarray}
with $E(S,T)$ the original system-time entanglement entropy \eqref{EST}, 
since $\rho'_S=\frac{1}{N}\sum_n|\psi_n\rangle\langle\psi_n|=\rho_S$,  Eq.\ \eqref{rhs} (in agreement with the general no signaling result that a local measurement at $T$ with no postselection cannot affect the reduced density of $S$),  while $\rho'_T=\frac{1}{N}\sum_n |n\rangle\langle n|$ is isospectral with $\rho'_{ST}$ 
and hence $S(\rho'_T)=S(\rho'_{ST})$, i.e.,  $S(S|T)=0$.  
The latter vanishes as the conditional states of $S$,  
$\langle n|\rho'_{ST}|n\rangle\propto |\psi_n\rangle\langle\psi_n|$,   
are pure $\forall\,|n\rangle$.

Hence, we can again measure the effective number of orthogonal  states visited in the now classically correlated mixed history state \eqref{rhopst} through essentially the same quantity $S(\rho_S)$, now equal to $I'(S,T)$. 

Remarkably, the same value of $I'(S,T)$ is obtained if the decoherence or local measurement takes place in {\it any} other orthogonal clock basis $\{|\tilde j\rangle\}$ of $T$,  $\langle \tilde j|\tilde j'\rangle=\delta_{jj'}$: 
The  history state \eqref{t0} can be rewritten in this clock basis as 
\begin{equation}
|\Psi\rangle= \sum_j \sqrt{p_j}\,|\phi_j\rangle\otimes |\tilde j\rangle,
\end{equation} 
where $|\phi_j\rangle=\sum_n \langle \tilde j|n\rangle|\psi_n\rangle/\sqrt{p_j}\propto \langle \tilde j|\Psi\rangle$ are normalized conditional system states  and $p_j=({\bf V}^\dag {\bf O}_N{\bf V})_{jj}\geq 0$, with ${\bf O}$  the  matrix \eqref{O} and ${\bf V}$ the unitary matrix of elements $V_{nj}=\langle n|\tilde j\rangle^*$ ($\sum_j p_j=1$).  Hence, after decoherence or a local measurement at the clock in this basis, we obtain 
\begin{equation}
\tilde\rho'_{ST}=\sum_j \tilde \Pi_j\rho_{ST}\tilde\Pi_j=\sum_j p_j |\phi_j\rangle\langle \phi_j|\otimes |\tilde j\rangle\langle \tilde j|\,,\label{rhopstj}
\end{equation}
where $\tilde \Pi_j=\mathbbm{1}_S\otimes|\tilde j\rangle\langle \tilde j|$.  Again $\tilde\rho'_S={\rm Tr}_T\,\tilde\rho'_{ST}=\sum_j p_j |\phi_j\rangle\langle\phi_j|=\rho_S$, Eq.\ \eqref{rhs}, in agreement with no signaling, 
while  $\tilde\rho'_{T}=\sum_j p_j|\tilde j\rangle\langle\tilde j|$ remains isospectral with $\tilde\rho'_{ST}$, such that   
$S(S|T)=0$ (as   $\langle \tilde j|\rho'_{ST}|\tilde j\rangle\propto |\phi_j\rangle\langle \phi_j|$ remains pure $\forall\,j$). Therefore, we still obtain  
\begin{eqnarray}
\tilde I'(S,T)&=&S(\tilde\rho'_S)+S(\tilde \rho'_T)-S(\tilde\rho'_{ST})\nonumber\\
&=&S(\rho_S)=E(S,T)\,.\label{Itpst}
\end{eqnarray}
In summary, $E(S,T)$ can also be considered as the {\it mutual information} of the decohered classically correlated mixed state which follows after an unread complete local measurement  at the clock in an arbitrary basis. 

In a classically correlated state like  \eqref{rhopst} or \eqref{rhopstj},  $I'(S,T)$ represents how much one can learn of system $S$ by observing or measuring the clock system $T$ (or viceversa). For an initially pure entangled state, this quantity is therefore independent of the choice of clock basis. This independence does not hold for classically correlated initial states, like the postmeasurement states \eqref{rhopst}-\eqref{rhopstj}.

\subsection{Generalized entropies  and majorization based characterization of evolution \label{IID}}
Other entropies can also be considered in Eq.\ \eqref{EST}. In particular, the R\'enyi entropies \cite{Beck.93} 
\begin{equation}S_q^R(\rho)=\frac{1}{1-q}\log {\rm Tr}\,\rho^q,\;\;\;q>0,\label{SR}\end{equation} approach the von Neumann entropy \eqref{S1}  for $q\rightarrow 1$ and attain  the {same previous lower and upper limits}  $\forall\,q>0$:   $S_q^R(\rho)=0$ for $\rho$  pure ($\rho^2=\rho$), $S_q^R(\rho)=\log N$ for $\rho$ maximally mixed ($\rho=\mathbbm{1}/N$, $N={\rm Tr}\,\mathbbm 1$), such that $0\leq S_q^R(\rho)\leq\log N$.\  The  ensuing    entanglement entropy is  
\begin{equation}
E_q(S,T)=S_q^R(\rho_S)=S_q^R(\rho_T)=\tfrac{1}{1-q}\log\sum_k\lambda_k^q\,,\label{EQR}
    \end{equation}
and the associated number of effective states visited is again $N_{E_q}=2^{E_q(S,T)}$ for $\log=\log_2$.  For $q\rightarrow 1$, $E_q(S,T)\rightarrow E(S,T)$. 

The  $q=2$ case, 
\begin{equation}
S_2^R(\rho)=-\log\,{\rm Tr}\,\rho^2\,,\label{R2}
\end{equation}
is particularly convenient  as it just depends on the purity ${\rm Tr}\,\rho^2\!$ and  does not require the calculation of the eigenvalues of $\rho$.
The associated entanglement entropy  can then be directly expressed 
in terms of the trace of the squared overlap matrix,   \begin{eqnarray}
E_2(S,T)&=&S_2^R(\rho_{T(S)})=
-\log {\rm Tr}\,{\bf O}_N^2\nonumber\\
&=&-\log {(}\tfrac{1}{N^2}\sum_{n, n'}|\langle \psi_n|\psi_{n'}\rangle|^2{)},\label{S2}
\end{eqnarray}
which admits an efficient evaluation 
($O(N^2)$ operations, instead of $O(N^3)$ operations when eigenvalues are required). 

Moreover, in a quantum realization of the history state,  efficient quantum protocols for direct evaluation of the purity of a density operator \cite{Naka.12}, and hence the entropy \eqref{S2}, and also the trace of an integer power of $\rho$ \cite{EAM.02,VED.18,Trace.25}, 
as required by the  R\'enyi entropies \eqref{EQR} for integer $q$, become available. These protocols  do not require a quantum state tomography.

Trace-form entropies \cite{PRL.02} $S_f(\rho)={\rm Tr}\,f(\rho)$, with  
$f:[0,1]\rightarrow\mathbb R$  a strictly concave function satisfying $f(0)=f(1)=0$ (and hence $f(\lambda)>0$ $\forall\,\lambda \in(0,1)$), could also be employed in \eqref{EST}. 
In particular,  for $f(\rho)=\frac{\rho-\rho^q}{q-1}$ we obtain the Tsallis entropies \cite{TS.09}  $S_q(\rho)=\frac{1}{q-1}(1-{\rm Tr}\,\rho^q)$, $q>0$, 
 which also reduce to the von Neumann entropy  for $q\rightarrow 1$ and are related to the R\'enyi entropies through  
$S^R_q=\frac{1}{1-q}\log[1-(q-1)S_q]$. The latter is just an increasing function of $S_q$ for $q>0$. 

These trace-form entropies are concave functions of $\rho$ and 
satisfy (under the previous conditions for $f$) the fundamental majorization inequality \cite{RC.02,Cap5Bet.13} \begin{equation}\rho\prec\rho'\Rightarrow S_f(\rho)\geq S_f(\rho')
\,,\label{prec}\end{equation} 
where $\rho\prec\rho'$ means $\rho$  {\it majorized} by $\rho'$ \cite{Bh.97,MOA.11} (see App.\ \ref{ApA}) and implies a condition involving all partial sums of the sorted eigenvalues of $\rho$ and $\rho'$ (Eqs.\ \eqref{prec00} and \eqref{prec11}).

Thus, majorization ensures  an {\it universal} 
entropy increase, being a more stringent criterion for mixedness or disorder than an  entropy increase for a single choice of entropic function. Moreover,  while $S_f(\rho)\geq S_f(\rho')$ for a single choice of $f$ does not imply majorization, the converse relation {\it does hold} if valid for {\it all} concave $f$: 
$S_f(\rho)\geq S_f(\rho')$ $\forall\,S_f$  $\Longleftrightarrow$  $\rho\prec\rho'$ \cite{RC.03}. Eq.\ \eqref{prec} is actually satisfied by all Schur-concave functions of $\rho$ \cite{Bh.97}, which include all R\'enyi  entropies \eqref{SR} for $q>0$.

When applied to the history state, 
Eq.\ \eqref{prec} allows us to identify three distinct evolution regimes when considering the system-time entanglement entropy  as a function of the number $N$ of times of the history. Denoting with $\rho^{(N)}_{S(T)}$ the reduced states of $S$ and $T$ in the history up to time $N$, we obtain: 

I. {\it Universal system-time entanglement entropy increase}: It arises if
\begin{equation}
\rho^{(N+1)}_{T(S)}\prec\rho^{(N)}_{T(S)}\,,\label{major}
\end{equation}
i.e., when $\rho^{(N+1)}_{S(T)}$ is {\it majorized} by $\rho^{(N)}_{S(T)}$. This ensures that
\begin{equation}
    E^{(N+1)}_f(S,T)\geq E^{(N)}_f(S,T)
    \label{maj}
\end{equation}
for {\it any} choice of entropy (i.e.,  of Schur-concave function of $\rho_{S(T)}$), including von Neumann, R\'enyi and all trace-form entanglement entropies   $E_f(S,T)=S_f(\rho_{S(T)})$. 

A related basic fundamental result is  the following: \\ {\it Lemma 1. Eq.\ \eqref{major} is satisfied 
whenever a  state $|\psi_n\rangle$ orthogonal to all previous  states, $\langle\psi_{n'}|\psi_n\rangle=0$ $\forall\,n'<n$,  is added to the history at step $n=N$ (sufficient  condition)}. The proof is provided at the end of App.\ \ref{ApA}.  

Therefore, by continuity, the
regime \eqref{major}--\eqref{maj} arises for some interval of values of $N$
essentially when the system visits a sufficiently distinguishable new state at each step of this interval.

II. {\it Universal system-time entanglement entropy decrease}: It arises if
\begin{equation}
\rho^{(N)}_{T(S)}\prec\rho^{(N+1)}_{T(S)}\,,\label{minr}
\end{equation}
i.e., when $\rho^{(N+1)}_{S(T)}$  {\it majorizes} $\rho^{(N)}_{S(T)}$. This implies 
\begin{equation}
    E^{(N+1)}_f(S,T)\leq E^{(N)}_f(S,T)
    \label{minj}
\end{equation}
for {\it any} choice of entropy. 

This is a very special case which requires, as necessary (but not sufficient) condition that the state added at the last step is already contained in the history, such that just a null eigenvalue  is added to the previous spectrum 
(otherwise the last majorization inequality cannot be fulfilled). 
In addition, the state added should be essentially among the most visited in the previous history, since otherwise it  could increase the weight of a seldom visited state, leveling the distribution and hence preventing this inverse majorization 
(see example and discussion at the end of App.\ \ref{ApA}).

III. {\it Non-universal regime:} Here Eq.\ \eqref{major} does not hold in any direction ($\rho^{(N+1)}\nprec\rho^{(N)}$ and $\rho^{(N)}\nprec\rho^{(N+1)}$), implying that some entanglement entropies will increase but others will decrease at the $N+1$ step. For example, one could stick to the von Neumann entropy 
and use just the associated entanglement entropy \eqref{EST}--\eqref{S1} as measure ($q\rightarrow 1$ in the R\'enyi case), but this does not ensure  that other choices of entropy 
 will exhibit the same behavior. 

  It should be also  noticed that conversion of an entangled pure bipartite state $|\Psi\rangle$ by LOCC (local operations and classical communication)  into another state $|\Phi\rangle$ requires majorization of the associated reduced densities (and not just $E(|\Psi\rangle)>E(|\Phi\rangle)$ for the von Neumann  or any other single choice of entropy) \cite{Nielsen.01},  so that Eq.\ \eqref{major} warrants a rigorous deep entanglement increase with increasing $N$ which is physically linked to the possibility of pure state conversion by LOCC to the lower $N$ history. Previous majorization based classification can obviously be also applied to compare any two different histories, like those for non consecutive times or different systems. 

 We finally mention that in the continuous limit, 
 the R\'enyi entropy of a gaussian distribution $N(\mu,\sigma^2)$ 
 is just 
 \begin{equation} 
S^R_q(\sigma)=C_q+\log\sigma\,,
\label{Sg}
\end{equation} 
exhibiting them all the same dependence with the standard deviation $\sigma$ except for a $\sigma$-independent constant 
$C_q=\tfrac{1}{2}[\log (2\pi)+\tfrac{1}{q-1}\log q]$. The result for the von Neumann entropy is recovered for $q\rightarrow 1$ ($\frac{1}{q-1}\log q\rightarrow \log e$).  A  similar result is valid for the  R\'enyi entropy of the exponential distribution $f(x)=\sigma^{-1}e^{-x/\sigma}$, $x\geq 0$, for which  $\langle x\rangle=\sigma=\sqrt{\langle x^2\rangle-\langle x\rangle^2}$ 
and \eqref{Sg} holds 
with $C_q=\frac{1}{q-1}\log q$. We have used  $S^R_q(f)=\frac{1}{1-q}\log\int_{-\infty}^\infty f^q(x)dx$ when applied to a continuous distribution $f(x)$. 
 
Thus, an approximate gaussian or exponential distribution of eigenvalues of ${\bf O}$  will lead roughly to  entanglement entropies $E_q$ differing in a $\sigma$-independent constant for distinct $q$, and exhibiting a similar behavior with $\sigma$. 

\subsection{Coherent states and overlap matrix\label{IIE}}

The well-known coherent states \cite{Milburn.07}  represent a family of  quantum harmonic oscillator states depending on a continuous variable, already introduced by E. Schr\"odinger \cite{schr.26}   
and later developed  by   R. Glauber in relation with  the quantum theory of coherence and quantum optics \cite{glauber1963coherent}, and also by E. Sudarshan \cite{sud.1963}. We will here use them to encode classical type data  based on real  numbers in quantum states, as discussed in the next subsection.  
Coherent  states 
possess minimum uncertainty (see below)  and can then be said to be the ``closest'' ones to classical states. 

Starting from bosonic creation and annihilation operators $a,a^\dag$, $[a,a^\dag]=1$, with vacuum state $|0\rangle$ ($a|0\rangle=0$), a coherent state $\ket{\alpha}$  is defined as the unique (except  for phase and normalization) eigenstate of the annihilation operator $a$ with eigenvalue $\alpha$: 
\begin{equation}
        a\ket{\alpha}=\alpha \ket{\alpha}\,. \label{eig}
\end{equation}
Such state exists $\forall\,\alpha\in\mathbb C$ and is just a displaced vacuum,  
\begin{subequations}
\begin{eqnarray}    \ket{\alpha}&=&D(\alpha)|0\rangle
=e^{-\frac{\abs{\alpha}^2}{2}}e^{\alpha a^\dag}|0\rangle\label{Dgen}\\&=&
e^{-\frac{\abs{\alpha}^2}{2}}\sum_{n=0}^{\infty}\frac{\alpha ^n}{\sqrt{n!}}\ket{n} \,,\label{coh0}
\end{eqnarray}
\label{coh}
\end{subequations}
where $D(\alpha)=e^{\alpha a^\dagger-\alpha^*a}$ is the unitary displacement operator ($D^\dag(\alpha) a D(\alpha)=a+\alpha$),   
    $\ket{n}= \frac{(a^{\dagger})^n}{\sqrt{n!}}|0\rangle$ are the Fock states ($a^\dag a|n\rangle=n|n\rangle$) and we used $e^{\alpha a^\dag+\beta a}=e^{\alpha\beta/2}e^{\alpha a^\dag}e^{\beta a}$.    
Thus, they lead to a Poisson distribution $|\langle n|\alpha\rangle|^2=\frac{|\alpha|^{2n}}{n!} e^{-|\alpha|^2}$  over the Fock states,  with $|\alpha|^2=\langle a^\dag a\rangle_\alpha$ the average boson number 
($\langle O\rangle_\alpha:=\langle \alpha|O|\alpha\rangle$). In quantum optics they are the states which best describe a laser beam \cite{Milburn.07}.

Besides, if $Q=\frac{a+a^\dag}{\sqrt{2}}$,  
$P=\frac{a-a^\dag}{\sqrt{2}i}$ 
 are the associated dimensionless coordinate and momentum ($[Q,P]=i\mathbbm 1$), and 
 $\Delta_Q^2=\langle Q^2\rangle -\langle Q\rangle^2$, $\Delta_P^2=\langle P^2\rangle-\langle P\rangle^2$, for averages in coherent states 
 we obtain $\langle Q\rangle_{\alpha}=\sqrt{2}\,{\rm Re}[\alpha]$, $\langle P\rangle_{\alpha}=\sqrt{2}\,{\rm Im}[\alpha]$ and $\Delta_{Q}^2=\Delta_P^2=\frac{1}{2}$,  
  hence leading to minimum uncertainty $\Delta Q\Delta P=1/2$.  We also recall that coherent states remain coherent when evolving in time under $H=\hbar\omega a^\dag a$: $e^{-iHt/\hbar}|\alpha\rangle=|\alpha(t)\rangle$, with $\alpha(t)=\alpha e^{-i\omega t}$, such that $\langle Q\rangle_{\alpha(t)}$, $\langle P\rangle_{\alpha(t)}$  follow the classical trajectory with  minimum uncertainty.

The coherent states fulfill  the completeness relation 
$\frac{1}{\pi}\int |\alpha\rangle\langle\alpha|d^2\alpha=\mathbbm{1}$ and are  nonorthogonal, having  a gaussian overlap 
\begin{equation}
\langle \alpha'\ket{\alpha}= e^{-\frac{1}{2}(|\alpha|^2+|\alpha'|^2-2\alpha{\alpha'}^*)}.\label{ov}
\end{equation}
In particular, for $\alpha,\alpha'$ real or having a common phase (i.e.\ $\alpha\rightarrow\alpha e^{-i\omega t}$ $\forall\,\alpha$, with $\alpha\in\mathbbm{R}$) 
this  overlap becomes 
\begin{equation}
\langle \alpha '\ket{\alpha}= e^{-\frac{1}{2}(\alpha-\alpha')^2},\label{ovr}
\end{equation}
corresponding to real gaussian wave functions $\langle q|\alpha\rangle=
\frac{1}{\sqrt[4]\pi}e^{-(q-\sqrt{2}\alpha)^2/2}$ (here  $Q|q\rangle=q|q\rangle$). 

Eq.\ \eqref{ovr} will be here of most importance. In first place it shows 
that any matrix with real gaussian elements $M^{(2)}_{\alpha\alpha'}=e^{-\frac{1}{2}(\alpha-\alpha')^2}$ 
is positive semidefinite for any set of real numbers $\alpha,\alpha'$, since it represents an overlap matrix of coherent states,  i.e., a Gram matrix (a matrix of inner products).  These hermitian matrices  are  positive semidefinite, and singular  only if the states involved are linearly dependent, their rank (number of nonzero eigenvalues) indicating the number of linearly independent states generated by the set considered. As any finite set of distinct coherent states is  linearly independent, the ensuing   overlap matrix  will  have  full rank unless some of the $\alpha$'s are strictly coincident. 
 
 It is worth noticing  that the exponent $2$ in the overlap function is the highest exponent yielding a positive semi-definite  matrix. In general, a matrix of elements 
 \begin{equation}
 M^{(p)}_{\alpha\alpha'}=e^{-|\alpha-\alpha'|^p/2}\end{equation} is not necessarily positive semidefinite for $p>2$, as can be checked numerically. 
 
 The gaussian case $M^{(2)}_{\alpha\alpha'}$ then lies at the ``border'' of the admissibility region for positivity.  
 The latter  can be explicitly verified from the Fourier identity $e^{-\alpha^2/2}=\frac{1}{\sqrt{2\pi}}\!\int_{-\infty}^\infty e^{i\alpha t} e^{-t^2/2}dt$, valid $\forall\,\alpha$. Hence  
 $\bm v^\dag M^{(2)} \bm v=\sum_{\alpha,\alpha'}v_{\alpha'}^*v_{\alpha}M^{(2)}_{\alpha'\alpha}=\frac{1}{\sqrt{2\pi}}\int_{-\infty}^\infty |v(t)|^2 e^{-t^2/2}dt \geq 0\,\forall$ vector $\bm v$ of arbitrary elements $v_\alpha\in\mathbb{C}$,  
 with $v(t)=\sum_\alpha e^{i\alpha t} v_\alpha$.  This integral  vanishes only if $v(t)=0\,\forall\,t$, entailing $\bm v=\bm 0$    for any finite set of distinct real numbers $\alpha$. 

We notice that 
a history of coherent states $|\alpha_n\rangle$, 
can be generated as in Eq.\ \eqref{gen}, using \eqref{Dgen}:
\begin{subequations}
\label{csh}
\begin{eqnarray}
    |\Psi\rangle&=&\tfrac{1}{\sqrt{N}}\sum_n|\alpha_n\rangle \otimes |n\rangle\label{HS0}\\
    &=&(\sum_n D(\alpha_n)\otimes |n\rangle\langle n|)(|0\rangle\otimes |\bar 0\rangle)\,,\label{Genc}\end{eqnarray}
        \end{subequations}
where $|0\rangle$ denotes the vacuum state of the boson 
and $|\bar 0\rangle=\frac{1}{\sqrt{N}}\sum_{n}|n\rangle=H_N|n=0\rangle$. 

We  also note that a pure Fock space representation of $|\Psi\rangle$ is feasible for e.g.\ a bosonic  clock, where states $|n\rangle$ could  be represented as  Fock states  $\frac{(b^\dag)^n}{\sqrt{n!}}|0\rangle$,  
with $b$ modes orthogonal to the $a$ modes
($[a,b^\dag]=[a^\dag,b]=0$, $[a,b]=[a^\dag,b^\dag]=0$), leading to $|\Psi\rangle=\frac{1}{\sqrt{N}}\sum_n D_a(\alpha_n)\frac{(b^\dag)^n}{\sqrt{n!}}|0\rangle$, 
with $|0\rangle$ the common vacuum  ($a|0\rangle=b|0\rangle=0$). We may also use other representations for the bosonic clock states $|n\rangle$, like well separated coherent states $|n\rangle=D_b(\beta_n)|0\rangle$ such that $\langle n|n'\rangle\approx 0$, or any other orthogonal set.  

We finally mention that we  could also use other states for encoding classical data, like  spin coherent states \cite{gilmore72,By.24}, which  can be easily realized with qubit systems and also admit a Fock-space representation \cite{By.24}.  Starting with 
a set of $L$ qubits and a single qubit state $\ket{\theta}=\cos\frac{\theta}{2}\ket{0}+\sin\frac{\theta}{2}\ket{1}$, real spin coherent states are  just aligned product states 
$\ket{\theta}^L:=\ket{\theta}^{\otimes L}=|\theta\rangle\otimes\ldots\otimes\ket{\theta}$: 
\begin{subequations}
\begin{align}
   \hspace*{-0.1cm} \ket{\theta}^L&=(\cos\tfrac{\theta}{2}\ket{0}+\sin\tfrac{\theta}{2}\ket{1})^{\otimes L}\\
    &=\sum_{M=-S}^{S}\!\! \sqrt{\textstyle\binom{2S}{S+M}}\cos^{S+M}\tfrac{\theta}{2}\sin^{S-M}\tfrac{\theta}{2}\ket{S,M},
    \label{sb}
    \end{align}
      \label{spcs}
   \end{subequations}
     $\!\!\!$where   \eqref{sb} is its total spin representation (identifying $|0\rangle$, $|1\rangle$ with the $\pm 1/2$ eigenstates of $s_z$), with $S=L/2$ the total spin and $|SM\rangle\propto S^{S-M}_-|0\rangle^L$  the eigenstates of the total $S^2$ and $S_z$ ($\bm S=\sum_{i=1}^L \bm s_i$), such that    $\langle\bm S\rangle_{\theta}=\frac{L}{2}(\sin\theta,0,\cos\theta)$. 
     
     The overlap between the states \eqref{spcs} is 
     \begin{equation}
          ^L\braket{\theta|\theta'}^L=\cos^L(\tfrac{\theta-\theta'}{2})
       \approx e^{-\frac{L}{2}(\frac{\theta-\theta'}{2})^2},   \label{spc}    
   \end{equation}
    where the last gaussian approximation holds for  sufficiently large $L$ and small $|\theta-\theta'|\lesssim \pi/2$. Notice that the exact expression in \eqref{spc} decreases with increasing $|\theta-\theta'|$ if $|\theta-\theta'|\leq\pi$. Since $|\theta\rangle=e^{-i\theta s_y}|0\rangle$, a history of states $|\theta_n\rangle^L$ can be constructed as in Eq.\ \eqref{csh}, with $|\alpha_n\rangle\rightarrow|\theta_n\rangle^L =D(\theta_n)|0\rangle^L$ and $D(\theta)=e^{-i\theta S_y}$ a tensor product of local rotations.  
  
\subsection{History state formalism for time series\label{IIF}}

In this section we present the application of   the history state formalism  to  time series. 
  In order to define the history state for a  time series of real numbers $p_n$, $n=0,\ldots, N-1$, we  assign to each $p_n$ a coherent state $|\alpha_n\rangle$ 
  of the form \eqref{coh}, 
  leading to a quantum encoding 
\begin{equation}|\alpha_n\rangle=|(p_n-\bar{p})/{\sigma}\rangle\,,\label{ali}
\end{equation}
where $\sigma\in\mathbb R$  is a scale determining the threshold of  distinguishability between two different values of the $p_n$'s (fixed for all   $p_n$  of a given history) and $\bar p$ is in principle an arbitrary real constant, which can be chosen to adapt the range of $\alpha_n$-values  to  a suitable interval.   
Since $\alpha_n\in\mathbb{R}$, the overlap between the  states \eqref{ali} becomes, using Eq.\ \eqref{ovr}, 
\begin{equation}    \langle\alpha_n|\alpha_m\rangle=
e^{-(p_n-p_m)^2/(2\sigma^2)}\,,\label{ov3}
\end{equation}
which is independent of $\bar p$. Hence $\langle\alpha_n|\alpha_m\rangle\approx 0$ when $|p_n-p_m|\gg \sigma$ while   
$\langle\alpha_n|\alpha_m\rangle\approx 1$ if $|p_n-p_m|\ll \sigma$. In the case of spin coherent states, 
 we may  just set  $\theta_n=\frac{p_n-\bar p}{\sigma\sqrt{L}/2}$ in \eqref{spcs} (choosing $\bar p$, $L$ in order that  $|\theta_n|\alt\frac{\pi}{2}$), such that for sufficiently large $L$ we end up with the same overlap \eqref{ov3}, according to Eq.\ \eqref{spc}.

Different choices of $\sigma$ will provide  more or less sensitivity to fluctuations in the values of the $p_n$'s: a small $\sigma$ will narrow the  gaussians, 
such that 
most states will be counted as distinct, leading to a larger entanglement entropy of the history state, whereas a large $\sigma$ will widen the gaussians,  hence counting many states as similar and yielding a smaller entropy. 
A standard criterion is to choose $\sigma$ of the order of some percentage of the range of values involved, so that a difference    $\abs{p_n-p_m}\approx \sigma$ in \eqref{ov3}   leads to an overlap $\langle\alpha_n|\alpha_m\rangle
\approx 0.6$, i.e.,  
a $\approx 40\%$ drop in comparison with its maximum, indicating that the associated values  start to be distinguishable.

While in the von Neumann case we need the eigenvalues of this overlap matrix for computing the entanglement entropy exactly, the quadratic R\'enyi entropy of Eqs.\ \eqref{R2}--\eqref{S2} can be directly evaluated: 
\begin{eqnarray} E_2(S,T)&=&-{\textstyle\log(\tfrac{1}{N^2}\sum\limits_{n, m}|\langle\alpha_n|\alpha_m\rangle|^2)}\label{E2x}\\
&\approx&\log N-{\textstyle\tfrac{2}{N}\!\!\sum\limits_{n<m}\!e^{-\frac{(p_n-p_m)^2}{\sigma^2}}}\,,
\label{E2a}\,\sigma\ll \Delta_{\rm min}\;\;\;\;\;\end{eqnarray}
where \eqref{E2a} is the approximation for small overlaps or equivalently  small $\sigma$ (and here $\log=\ln$) such that  $\sum_{m<n}|\langle \alpha_n|\alpha_m\rangle|^2\ll N$, i.e.\ almost orthogonal states ($\Delta_{\rm min}$ denotes the minimum absolute difference). 

On the other hand, if the aim is to measure just  significant variations, one could choose a sufficiently large $\sigma$ 
such that $(p_n-p_m)^2/\sigma^2\ll 1$, and hence,   \begin{equation}\langle\alpha_n|\alpha_m\rangle\approx 1-\tfrac{(p_n-p_m)^2}{2\sigma^2}\,.\label{ovl}\end{equation}
In this limit, opposite to that of \eqref{E2a}, Eq.\ \eqref{E2x} becomes 
\begin{subequations}
\label{E2l}
\begin{eqnarray} E_2(S,T)&\approx&-\log(1-\!\!\tfrac{2}{N^2\sigma^2}{\textstyle\sum\limits_{n<m}
(p_n-p_m)^2})\\&\approx&\tfrac{2}{N^2\sigma^2}{\textstyle\sum\limits_{n<m }(p_n-p_m)^2}\label{s2ap1}\\&=&\tfrac{2}{\sigma^2}[
 \tfrac{1}{N}\!\sum_n p_n^2\!-\!(\tfrac{1}{N}\!\sum_n p_n)^2],\;\sigma\gg \Delta_{\rm max}\;\;\;\;\;\;\;\;\;\label{s2ap2}
\end{eqnarray}
\end{subequations}
which is just proportional to the {\it fluctuation}, i.e.\ the variance of the   $p_n$'s in the history considered ($\Delta_{\rm max}$ is the maximum price difference, and again  $\log=\ln$). 
Moreover, for  $p_n>0\,\forall\,n$, one could also use the  {\it logarithm}  of the values $p_n$ in order to measure just very significant variations, i.e., $p_n\rightarrow \log p_n$ in Eqs.\ \eqref{ali} and \eqref{ovl}--\eqref{E2l}.

\section{\label{III}Application to financial time series}

\begin{figure}[h]
\centering
    \includegraphics[width=.8\linewidth]{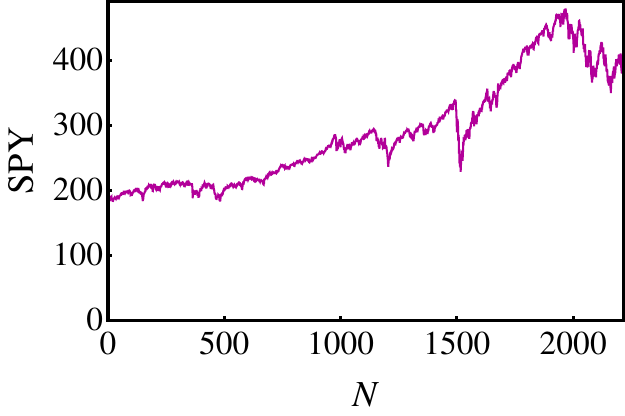}
    \vspace*{-.25cm}
    
\caption{Opening values of the SPY (in USD) as a function of the elapsed number of days $N$ along the period March 17, 2014 -- December 30, 2022 
\cite{Note1}.}
    \label{fig1}
\end{figure}
\vspace*{-.0cm}

\subsection{Full history analysis}\label{III.A}

We now apply this formalism to 
 the analysis of  the  evolution of financial time series. 
We first consider the SPY, the  code for the ETF
which tracks the SPX stock market index, in the period from March 17, 2014 to December 30, 2022,  comprising 2206 trading days. We  employ the daily opening price $p_n$ as the representative price of that day. They are  
 shown in Fig.\ \ref{fig1} with data taken from \footnote{
 Yahoo Finance, `` SPDR S{\&}P 500 ETF Trust (SPY) Historical Data,'' 
http://finance.yahoo.com/quote/SPY/ history/(2024)}. The data for all Figs.\ can be found in \footnote{F. Lomoc, N. Canosa, A.P. Boette, R. Rossignoli, ``QF supplemental data'',  https://github.com/ferlamot/QF-supplemental-data (2026)}.

We first consider the history state associated to previous prices for $N$ contiguous trading days,  starting on  March 17, 2014, for $1\leq N\leq 2206$. 
We use the coherent state embedding \eqref{ali}, 
hence with overlaps \eqref{ov3}.

\subsubsection{Entanglement entropy}
\begin{figure}[h]
\begin{center}
\includegraphics[width=.95\linewidth]
{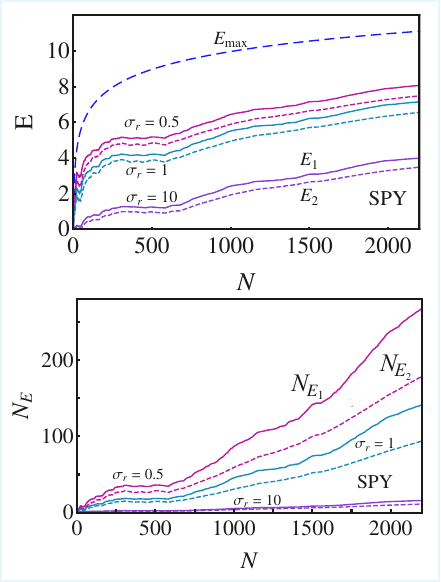}
\end{center}
\vspace*{-.75cm}

\caption{Top panel: The von Neumann (solid lines) and quadratic R\'enyi (short dashed lines) entanglement entropies $E_1$ and $E_2$ respectively,  Eqs.\ \eqref{EST2}--\eqref{S2},
of the history state associated to the SPY prices for the first $N$ days, $1\leq N\leq 2206$, starting on March 2014, for distinct values of the relative width $\sigma_r$.   
For reference the maximum attainable value $E_{\rm max}=\log_2 N$ (the same for  both entropies, representing the $\sigma_r\rightarrow 0$ limit),  is also depicted. Bottom: The corresponding effective number  of distinct states  visited in the same period, Eq.\ \eqref{NS}, 
  for the indicated entropies, as a function of $N$.}
\label{fig2}
\end{figure}

  We depict in Fig.\ \ref{fig2} the entanglement entropy of the history state 
  up to day $N$ for the whole period,  for different values of $\sigma$ in  \eqref{ali}--\eqref{ov3}.  We have set  $\sigma=\sigma_r \sigma_0$,  with $\sigma_0=1$  USD taken as unit and $\sigma_r$ a relative dimensionless width.  We show results for the two entropic measures that will be employed in what follows: the  von Neumann  entropy 
  \eqref{EST2}, denoted as $E_1$  in the plots, and  the quadratic R\'enyi  entropy \eqref{S2},   denoted as $E_{2}$. 
 
  Both entanglement entropies, which measure essentially the logarithm of the effective number of distinct prices visited in the history, evolve similarly  except for an almost constant vertical shift $\Delta E\approx 0.5$, for each fixed value of $\sigma_r$, though $E_1$ exhibits a slightly higher sensitivity. They both display a rather smooth steady 
  increase for increasing $N$ if $N\agt N_c\approx 585$,   reflecting the upward trend of prices,  as new values are ``visited''. Nonetheless, they also show some small slope discontinuities at certain days, with short intervals of negative slope for $N<N_c$,  which indicate essentially that no distinct states (i.e.\ prices)  are visited at those days.

  Remarkably, the behavior for distinct widths $\sigma_r$ is also similar in the indicated range, except for an almost constant difference which depends on the ratio of the $\sigma_r$: The entanglement entropy of the history obviously increases for decreasing $\sigma_r$, i.e., increased resolution. Roughly, 
  $E_{\sigma_r=0.5}-E_{\sigma_r=1}\approx 0.9$ for $N\agt 300$, while 
   $E_{\sigma_r=1}-E_{\sigma_r=10}\approx 3.1$, in agreement with a dependence similar to Eq.\ \eqref{Sg}.  
 We also show for comparison the  common maximum entropy $E_{\rm max}=\log N$ $\forall\,q>0$   that can  be obtained up to that day,  which  corresponds essentially to the $\sigma_r\rightarrow 0$ limit and would be   reached only if a new distinct state, i.e., a well distinguishable asset price not contained in previous history, were reached at each day.  For the chosen values of $\sigma_r$, $E_{\rm max}$ is   much higher than the actual history entropies, indicating  a price behavior well detached from the saturation limit.  
 
 In the bottom panel we show  the effective number \eqref{NS} of distinct states $|\alpha_n\rangle$  visited during the evolution, for the von Neumann  and  quadratic R\'enyi entropies.  It is verified that $N_E$ lies well below $N$.    
 The behavior for both entropies is again quite similar, since their ratio for fixed $\sigma_r$ and distinct entropy, as well as for fixed entropy and distinct $\sigma_r$, is almost constant due to the roughly constant entropy difference shown in the upper panel. Again,   $N_{E}$   shows a slightly higher sensitivity in the von Neumann case. The slope variations, indicative of changes  in the price evolution regime,   and the intervals of entropy decrease,  are more clearly appreciable. 
 
 \begin{figure}[h]
\begin{center}
    \includegraphics[width=0.95\linewidth]
       {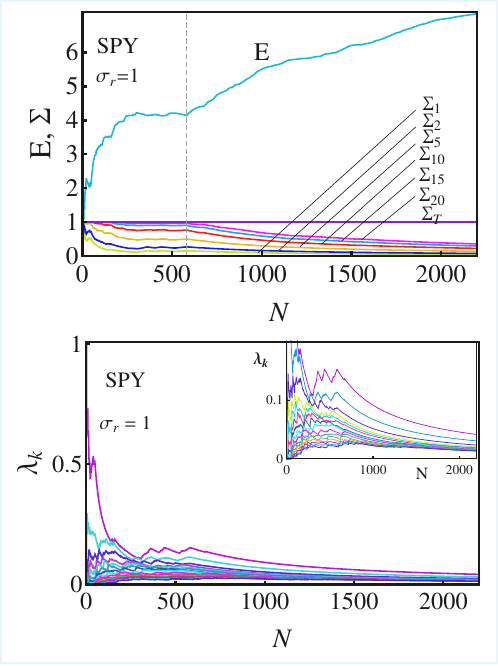}
\end{center}
\vspace*{-.65cm}

\caption{Top panel:
The entanglement entropy $E=E_1$ of the history state 
as a function of the number of elapsed days $N$  for $\sigma_r=1$  in the whole period, and 
the associated partial sums $\Sigma_{\ell}=\sum_{k=1}^{\ell}\lambda_k$ of the eigenvalues of the overlap matrix, as a function of $N$,  for  the indicated values of $\ell$, with  $\Sigma_{T}=1$ the total sum. 
The vertical dashed line signals a regime change after which all partial sums tend to decrease, indicating  majorization for increasing $N$ 
after this point.  
Bottom panel: The corresponding  entanglement spectrum of the history state 
as a function of the number of elapsed days $N$. We depict the first 15 eigenvalues $\lambda_k$ of the overlap matrix {\bf O}$_N$, (which are those of $\rho_S$ and $\rho_T$), with the elements \eqref{ov3}. 
The inset shows in more detail the behavior of the lowest eigenvalues.}
\vspace*{-0.65cm}
\label{fig3}
\end{figure}

\subsubsection{Entanglement spectrum and majorization}

In order to obtain a deeper  picture of the  entanglement entropy evolution, 
we now analyze 
the entanglement spectrum of the history state  and the associated majorization properties.

In the top panel of Fig.\ \ref{fig3} we show the evolution of the von Neumann entanglement entropy $E=E_1$  for $\sigma_r=1$, together with the corresponding partial sums  
\begin{equation} 
\Sigma^{(N)}_{\ell} =\sum_{k=1}^{\ell}\lambda_k^{(N)}
\label{suml} 
\end{equation}
of  the eigenvalues $\lambda_k^{(N)}$ of the normalized overlap matrix for the first $N$ steps, which determine their majorization properties (section \ref{IID} and App.\  A). 
It is seen that  it is possible to clearly identify two different regimes during the evolution. For $N\agt N_c=585$ (vertical dashed line) all  partial sums \eqref{suml} after $N$ steps  decrease with increasing $N$, entailing that   majorization relations \eqref{prec}--\eqref{prec1} are fulfilled and hence the universal entropy increase regime I is taking place, i.e.,  not only the depicted von Neumann entropy, but all entanglement entropies, are increasing. This behavior indicates  that  
essentially ``new'' prices are incorporated to the history at each step. 

In contrast,  in the previous sector $N\alt N_c$ there are values or even intervals of $N$ for which the partial sums are not majorized by those  of  previous $N$ and this is reflected in the  rather stable value of the entanglement entropy for $\tfrac{1}{2}N_c\alt N\alt N_c$, above the initial increase.  In this intermediate region  
no strict majorization relation takes place between the spectra for different $N$ (type III   regime). This is in accordance with the ``lateral phase" (no clear trend), in which prices are confined to a range.    
Besides,  at certain values of $N$  the partial sums are approximately majorized by those of the next step, leading to an almost universal slight entropy decrease at these points 
(type II regime). 
The vertical line at $N=N_c$ then signals a noticeable change in the behavior of  the partial sums and the associated entropy, which enables to identify different regimes. 
 
 In the bottom panel of Fig.\ \ref{fig3}  we show the evolution with $N$ of the entanglement spectrum of the  normalized overlap matrix {\bf O}$_N$ of  Eq.\ \eqref{O}, with the elements \eqref{ov3}.  
 It seen that in the sector $N\agt N_c$ the largest eigenvalues exhibit a rather continuous decrease with increasing $N$, reflecting the incorporation of new small eigenvalues, i.e., orthogonal states in the history, in agreement with  the universal entropy increase regime I. There is here a progressive increment in the number of  small eigenvalues whose  magnitudes initially  increase. 
   In contrast, in the sector $N\alt N_c$ above the initial increase, the eigenvalues exhibit noticeable oscillations (see also inset), leading to a different regime in the partial sums (type III, no strict majorization) and entropy, except for short intervals  of almost type II behavior. 

\subsection{Evolution of monthly histories}\label{III.B}
 It is also possible to use the formalism for   other  types of analyses of  the time series along  a given period, by  considering partial histories, i.e., histories along a fixed shorter time interval. 
 In this way the ensuing entanglement entropy will be a measure of the  effective number of distinct  prices reached in such interval, providing an indicator of  the  volatility of the asset in the latter.  The evolution of such entropies within the total period   will then provide a picture  of the volatility variation.

In this section 
we consider the evolution of the SPY asset along a period of one month, corresponding to at most 21 trading days.  
Accordingly, we construct  the  history state for each month and evaluate the associated entanglement entropy and spectrum.  We then examine these quantities for a total period of 120 months, from March 19, 2014 to March 15, 2024 (120 trading months). 

\begin{figure}[h]
\centering    \includegraphics[width=0.95\linewidth]
{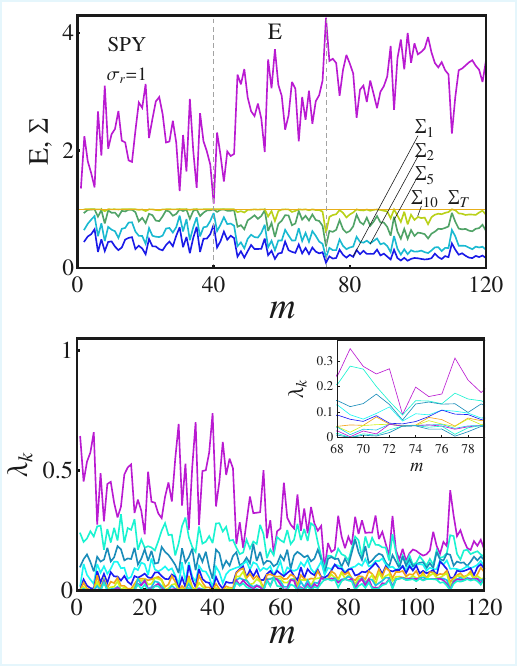}
\vspace*{-0.5cm}
\caption{ Top: The entanglement entropies $E=E_1$ of the one-month  history states for  120 months $m$ starting on March 17, 2014, for $\sigma_r=1$,  together with the associated   partial sums $\Sigma_{\ell}=\sum_{i=1}^{\ell}\lambda_k$ of the entanglement spectrum for each month for $\ell=1,2,5,10$.  The vertical dashed lines indicate  months 73 and 40, where the monthly entropy reaches its maximum and minimum respectively, and accordingly, the monthly partial sums have their minimum and maximum.  
Bottom: The corresponding entanglement spectrum of the monthly histories in the same period. We depict the first 10 eigenvalues $\lambda_k$ of the normalized overlap matrix for each  month. The inset shows in more detail the behavior of the lowest eigenvalues around the region of highest monthly volatility ($68\alt m\alt 79$).}
    \label{fig4}
    \vspace*{-0.75cm}
\end{figure}

 In the top panel of Fig.\ \ref{fig4} we show results for the von Neumann  entanglement entropy $E=E_1$ of the history of each month together with the corresponding  partial sums of the eigenvalues of the normalized overlap matrix ${\bf O}_{N_m}$ ($N_m\leq 21$)  for the sequence of 120 months of the period of Figs.\ \ref{fig2} and \ref{fig3} with $\sigma_r=1$. The vertical dashed line at month 73 indicates that at this  month all partial sums are lower than those of any other month in the whole period, implying that the spectrum of this month is majorized by that of any other month.  This  entails a  maximum monthly entanglement entropy at this month, for {\it any} choice of entropy (and not just for the depicted von Neumann case),  indicating the month with a maximum number of distinct asset prices for the $\sigma$ considered, i.e., of maximum volatility.   
On the other hand, the vertical dashed line at month 40  indicates the month of minimum entropy, where the partial sums are all greater  than those of any other month and hence the spectrum majorizes that of any other month. Hence, it indicates the month of minimum price dispersion, i.e., minimum volatility. 

In the bottom panel of Fig. \ref{fig4}, we show the evolution of the  10 largest eigenvalues of the monthly history  entanglement spectrum, along the whole period. Maximum (minimum) entropy corresponds obviously to minimum (maximum) dispersion of eigenvalues. From the associated Schmidt decomposition, we can say that these eigenvalues represent the normalized probability distribution of the actual  distinct prices (according to the chosen $\sigma_r$) in the corresponding month. 
The inset depicts the detailed behavior of the main eigenvalues in the vicinity of the month of maximum entropy (i.e., maximum volatility). At the maximum they tend to be almost degenerate, having minimum dispersion.

\begin{figure}[h]
\begin{center}
\includegraphics[scale=0.7]
{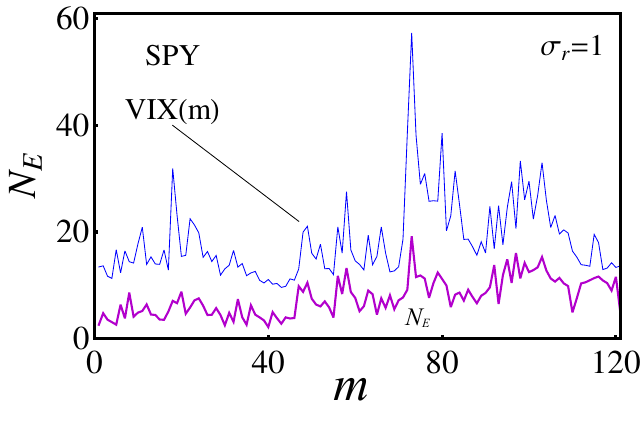}
\end{center}
\vspace{-.75cm}
\caption{Number of effective visited  states $N_E$ for each month, in the same case of Fig.\  \ref{fig4}.  
The monthly VIX values for the same period are also depicted. There is a close qualitative agreement between both quantities.}
\label{fig5}
\end{figure}

In Fig.\ \ref{fig5}, we show the evolution of the corresponding effective number \eqref{NS} of monthly visited ``states'' $N_E$ for the von Neumann  entropy,  and compare it with the monthly VIX index values in the same period, taken from 
\footnote{Yahoo Finance,``CBOE Volatility Index (VIX) Historical Data,''http://finance.yahoo.com/quote/{\%}5EVIX/history\\/(2024).}. 
 There is clearly a good qualitative agreement between both quantities,  with the maxima of  $N_E$ matching those of the VIX. 
Accordingly,  the maximum of the monthly values of $N_E$ signals the largest peak  of the monthly VIX values, located at month 73.  Also, the minimum  $N_E$, at month 40, lies close to the minimum of the VIX. 
Hence, $N_E$ can be considered as an alternative measure of volatility.

\vspace*{-1cm}

\subsection{Evolution of weekly histories}\label{III.C}

In order to examine in detail the behavior around a specific point, in this section we  study the SPY histories for just one week, comprising at most 5 consecutive trading days, and analyze their evolution in a surrounding interval.  We will focus on the interval comprising the maximum weekly volatility (which  corresponds essentially to the months of maximum monthly volatility),  where the weekly VIX index experiences its largest variation. Namely,   
an interval of 60 weeks from week 300 up to week 360 in the whole previous period of 522 weeks  from March 17, 2014 to March 15, 2024.

\begin{figure}[h]
\begin{center}
\hspace*{-0.2cm}\includegraphics[width=1.05\linewidth]
{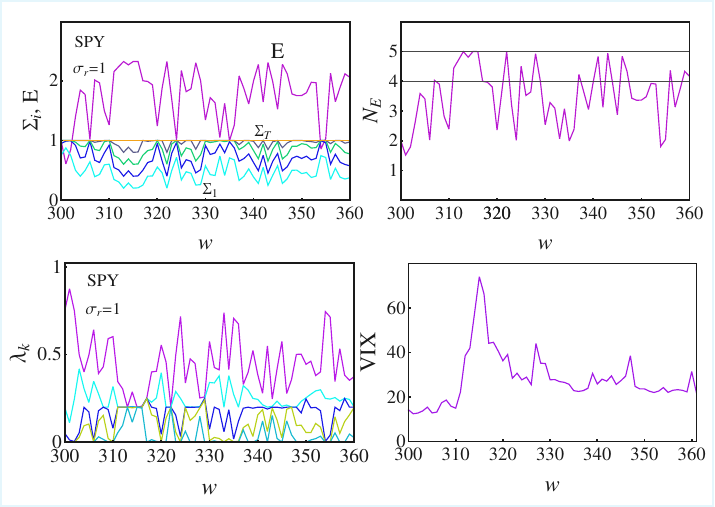}
\end{center}
\vspace{-.5cm}
\caption{
Top left: The von Neumann entanglement entropies $E=E_1$ of the one-week history states for the indicated 60 weeks period,  together  with the associated partial sums $\Sigma_\ell$ ($\ell=1,\ldots,5=T$) of the   entanglement spectrum for each week. 
Top right: The corresponding effective   number of weekly visited states $N_E$, Eq.\ \eqref{NS}, in the same period. 
Bottom: Evolution of the weekly entanglement spectrum (left panel) and weekly VIX (right panel) in the same period.  It is seen that the VIX 
experiences a noticeable variation  in  the weeks with highest occupation $N_E$  and highest spectrum degeneracy, taking its maximum where there is full degeneracy and $N_E$ is maximum.} 
\vspace*{-0.5cm}
\label{fig6}
\end{figure}
 
In the top left panel of Fig.\ \ref{fig6} we show the evolution of the weekly von Neumann entanglement entropy  $E=E_1$ in this interval, together  with the corresponding  partial sums $\Sigma_\ell$  for $\ell=1,\ldots,5$ ($N_w\leq 5$).  
At  weeks 313 and 315, the partial sums are minimum, i.e., lower than those of any other week,  and accordingly, the entropy is maximum, reaching saturation $E=\log N_w$, reflecting a plain uniform distribution. 
 Majorization occurs and any entropy will then be maximum at these two weeks. 
 
The top right panel  shows the  effective number of visited states $N_E$  in the corresponding week. It obviously saturates at the weeks of maximum entropy, signalling  maximum weekly volatility.   
On the left bottom panel we show the  detailed evolution of the set of  eigenvalues (entanglement spectrum) of ${\bf O}_{N_w}$ for each week along the same period. The  presence of complete or almost  complete degeneracy   in the spectrum  can be clearly identified at  weeks 313, 315, and some other values (like 322, 329, etc.), 
which coincide with the highest maxima of $N_E$ ($N_E>4$) of the top right panel, hence indicating high volatility. In fact, 
weeks for which the spectrum is fully or partially degenerate correspond approximately to those weeks where maxima or abrupt changes in the VIX values (bottom right panel) occur. 
This weekly analysis then enables to rapidly recognize  periods of high volatility of the asset. 

\subsection{Weekly and monthly logarithmic histories}\label{III.D}

In order to explore other approaches feasible within the history state-based formalism, and with the aim of 
obtaining an improved resolution of 
 the most prominent  fluctuations above the baseline noise, in this section we analyze the histories of the {\it logarithm} of the prices. 
In this case we set 
$\alpha_n=\log [p_n/\bar p]/\sigma$ in the quantum coherent state representation, where $\bar p$ is some price scale. 
Then, according to Eq.\ \eqref{ovr}, their overlaps become  
$\langle \alpha_n|\alpha_m\rangle=
e^{-\log^2[p_n/p_m]/(2\sigma^2)}$, again  $\bar p$ independent and determined just by the logarithm of the prices ratios. 

Since such logarithms 
 are relatively small, this approach will essentially pick up  the largest price variations. Moreover, already for not too small values of $\sigma$ (i.e.\ $\sigma\agt 1$), we are already in the large overlap regime of Eq.\ \eqref{ovl},  such that the $q=2$ R\'enyi  entanglement entropy of the associated history state 
approaches Eqs.\ \eqref{E2l}. It then  becomes  proportional to $\sigma^{-2}$ times the {\it fluctuation} of the logarithm of the asset price along the $N$-step history, 
according to Eq.\ \eqref{s2ap2}, and hence to 
$\frac{1}{N^2}\sum_{m<n}\log^2(p_n/p_m)$ according to   \eqref{s2ap1}, leading to  
\begin{equation}
    E_{2_{\rm log}}(S,T)\approx \tfrac{2}{N^2\sigma^2}\sum_{n<m}\log^2(p_n/p_m)\,.\label{E2log}
\end{equation}
In the following we  then employ the quadratic R\'enyi entropy with $\sigma=1$, 
such that its value is essentially the logarithmic fluctuation of the prices, Eq.\  \eqref{E2log}, except for a scale factor $\propto \sigma^{-2}$. 

\subsubsection{SPY Histories}\label{III.D1}
We first apply this scheme to  further analyze  the evolution of  the SPY along the whole previous period of ten years, based on the variation of weekly and monthly logarithmic history entropies (Eq.\ \eqref{E2log}). For reference we compare results with those of the corresponding VIX for the SPY. 

\begin{figure}[h]
{\centering
\includegraphics[width=.95\linewidth]
{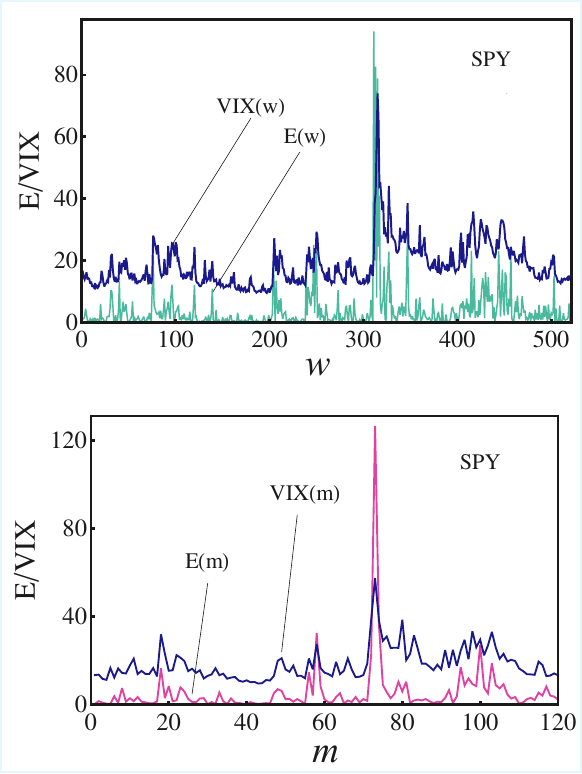}}
\vspace{-.25cm}
\caption{ 
Evolution of the quadratic R\'enyi entanglement entropy of the weekly (top) and monthly (bottom) histories, employing the logarithm of SPY prices (see text), together with the corresponding reference VIX values  for  SPY, along the whole previous period. 
}
\label{fig7}
\end{figure}  

It is worth mentioning that the VIX needs the input of several options prices in a range of around 30 days, and can be cumbersome to calculate for a type of asset different from the SPY. Here we  examine the ability of  the previous logarithmic entropic measure of volatility  to account for the main features of VIX index evolution. 

In  Fig.\ \ref{fig7} we depict the evolution of the weekly (top) and monthly (bottom) VIX index for a period  of ten years, corresponding to 522 weeks and 120 months,
together with the corresponding entropic measure of volatility given by the $q=2$ R\'enyi entropy  $E_2$, Eq.\ \eqref{S2},   using the logarithm of prices, which leads essentially to Eq.\ \eqref{E2log}.   
For a better visualization and comparison with the VIX in the same plot, we have scaled our results  by multiplying them by a fixed factor (equivalent to a  fixed value of $\sigma$ in the approximation \eqref{E2log}). 

It is seen that our  measure  is able 
to reproduce the main features of the corresponding VIX values, 
with the peaks of the entropic measure matching the most relevant changes in the VIX evolution and distinguishing the main peaks (points of large volatility in the asset), 
for both weekly and monthly evolutions. The use of the logarithm tends to filter the noise, providing  an improved signal-to-noise ratio,  being more sensitive to the higher price fluctuations. 

\subsubsection{Results for Gold and Bitcoin}\label{III.D2}

\begin{figure}[ht]
{\centering
\includegraphics[width=.9\linewidth]
{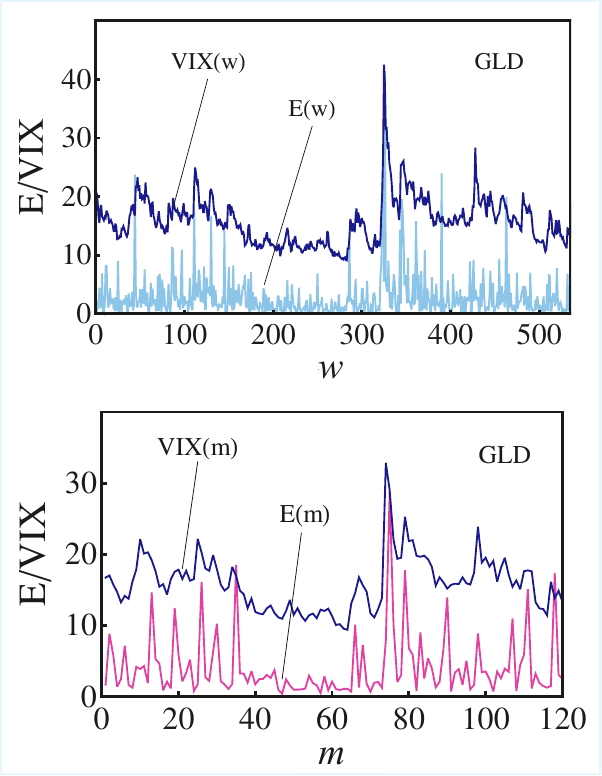}}
\vspace*{-.25cm}
\caption{Top: Evolution of the  quadratic R\'enyi  entanglement entropy of weekly histories, employing the logarithm of Gold prices  
together with the corresponding weekly VIX values, along a period of  535 weeks. Bottom: The same quantities for monthly histories and VIX values, for a period of 120 months.}
    \label{fig8}
\end{figure}

\begin{figure}[h]
\begin{center}
\includegraphics[width=.9\linewidth]{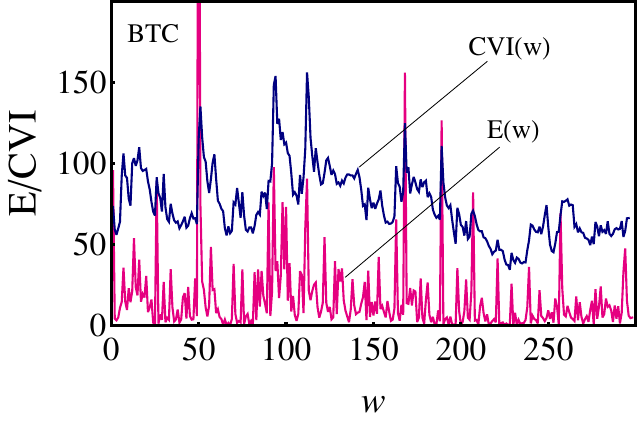}
\end{center}
\vspace*{-.75cm}
\caption{Evolution of the  quadratic R\'enyi entanglement entropy of weekly histories, employing the logarithm of BTC prices, and the reference CVI values for BTC, along a  298 weeks period.}
\label{fig9}
\end{figure}

Finally, we also include results for
GLD  \footnote{Yahoo Finance, ``SPDR Gold Shares (GLD) Historical Data'' https://finance.yahoo.com/quote/GLD/history}  
 and BTC \footnote{Yahoo Finance, ``Bitcoin (BTC-USD) Historical Data,''
http://finance.yahoo.com/quote/BTC-USD/history/(2024)}. 
Comparisons between  the present entropic measure of volatility  and  the standard volatility measures for these assets 
(CBOE Gold ETF Volatility Index for GLD 
\footnote{Federal Reserve Bank of St. Louis, ``CBOE Gold ETF Volatility Index,''http://fred.stlouisfed.org/series/\\GVZCLS/(2024)} 
 and CVI for BTC \footnote{Investing.com,``Crypto Volatility Index (CVI),'' http://es.investing.com/indices/crypto-volatility-index-historical-data/(2025)} 
  are shown  in Fig.\  \ref{fig8} for gold,  for the same previous period of 10 years,  and 
  in Fig.\ \ref{fig9} for BTC, for a  298 weeks period.  
   
In the top panel of Fig.\ \ref{fig8}, the results correspond to a period of 535 weeks, where the  overlaps were evaluated for periods of  one week (5 days), whereas in the bottom panel results are for the same interval corresponding to 120 months, with the  overlaps  evaluated for periods of  one month (21 days).  In both cases, we have employed the entropic measure of volatility given by the $q=2$ history entanglement R\'enyi  entropy,   
employing the logarithm of the GLD prices, which again leads essentially to Eq.\ \eqref{E2log}. We have also employed a convenient fixed scale in order to compare our results with those of  the VIX index.  

In Fig. \ref{fig9} we show the evolution of  the   CVI  index for BTC, together with the evolution of our entropic measure of volatility as a function of time in weeks. The latter  was again evaluated with the $q=2$ R\'enyi entanglement entropy $E_2$,  and we have employed again the logarithm of the BTC prices (leading to Eq.\ \eqref{E2log}).   We have    analyzed a period of  298 weeks, from March 31, 2019 to  December 20, 2024.

As in the case of the SPY, results  for these two different assets  show that the entropic measure of volatility  is able to clearly identify those  weeks or months 
which are characterized by a large volatility of the asset. There is a 
 good agreement  between results from our approach and those of the standard measures of volatility corresponding to the evolution of each asset.
We see that in general, for the analyzed assets, our entropic measure has a better signal-to-noise ratio than the standard volatility indicators when using the logarithm of prices, as peaks are more prominent than the baseline volatility 
as compared to the VIX and the CVI. 

\section{Conclusions} \label{IV}

We have introduced a general method  for  analyzing  time series from the new perspective provided by the quantum history state formalism. This framework leads to the concept of system-time entanglement,  which constitutes a measure of the  effective number of orthogonal (i.e., fully distinguishable) states visited during the whole evolution considered. It is fully determined by the eigenvalues of the overlap matrix of the evolved states, which here form  the  entanglement spectrum.  It also emerges as mutual information  of the decohered mixed history state. 

The  history state associated to a given time series is then  defined through a scaled quantum coherent state embedding of the time series data. Through the corresponding gaussian overlaps between these  states, a consistent smooth transition between coincident and clearly distinguishable data is achieved.  The associated entanglement entropy then provides a measure of the effective number of distinct data in the series, relative to the reference width. The ensuing entanglement spectrum represents the relative frequency of appearance of the effective ``orthogonal'' states of the evolution, the latter emerging  from the Schmidt decomposition of the history state. 
In addition, a rigorous  analysis of the entanglement spectrum based on majorization theory allows one to identify distinct evolution regimes, including those characterized by a universal entanglement entropy increase or decrease. 

The formalism was used to examine financial time series, starting with  the price evolution of the SPY asset. Through different types of histories, including the evolution of the SPY along a full history in \ref{III.A},  the evolution of monthly histories in \ref{III.B}, and  the evolution of weekly histories in  \ref{III.C}, it was seen that the corresponding  entanglement spectrum, entropy and associated effective number of visited states $N_E$, provide reliable indicators of relevant properties of the  series. 
In particular, from the majorization properties of the  spectrum, different evolution regimes in the full history of the SPY could be identified,  whereas 
through monthly and weekly histories, a good correspondence 
between the entanglement entropy (or  the ensuing $N_E$)  with the pertinent VIX values was achieved,  thus detecting periods of high volatility. 

In order to obtain an improved visualization of main variations, in \ref{III.D} we  also explored weekly and monthly logarithmic histories and entanglement entropies, for SPY as well as for GLD and BTC, and compared them with the corresponding  volatility  measures. 
A close qualitative agreement between them was obtained, as highlighted from the matching in the position of the main peaks observed in the figures. Moreover, the quadratic R\'enyi entropy of the logarithmic history was shown  to be closely related to the logarithmic fluctuation. 

In summary, the present quantum-based formalism is  able to provide deep insights into the complex behavior of  time series, including  measures   showing a good agreement with standard volatility indicators of financial series.   
 This quantum formulation also opens the way to  other possibilities,  from quantum simulation of the full history, which could enable the implementation of quantum computation protocols for efficient evaluation of time averages \cite{BRGC.16,DN.25} and entanglement entropies through trace estimation \cite{Naka.12,EAM.02,VED.18,Trace.25}, to possible new approaches for  describing the evolution of time series, 
which are currently under investigation. 

\acknowledgments 
Authors acknowledge support from CONICET (F.L., N.C.) and CIC (R.R.) of Argentina.  Work supported by
CONICET PIP Grant No. 11220200101877CO.
\appendix

\section{Majorization and mixedness. }
\label{ApA}

Let ${\bf p}=(p_1, p_2,...,p_n)$ and ${\bf q}=(q_1, q_2,..., q_n)$ be two probability distributions ($p_i\geq 0$, $\sum_i p_i=1$ and similarly for ${\bf q}$). 
If we sort the components of ${\bf p}$ and ${\bf q}$ in decreasing order, such that $p_1\geq p_2\geq...\geq p_n$, $q_1\geq q_2\geq...\geq q_n$, it is said that ${\bf q}$ is {\it majorized} by ${\bf p}$ \cite{Bh.97,MOA.11}
 or, in symbols, 
 \begin{subequations}\begin{equation}
{\bf q} \prec {\bf p} \,,
\end{equation} 
iff the following inequalities are all fulfilled: 
\begin{equation}
\sum_{j=1}^i{q_j}\leq \sum_{j=1}^i{p_j}\;\;\forall \, i=1,\ldots,n-1\,, \label{psums}
\end{equation}
\label{prec00}
\end{subequations}
$\!\!$with equality for $i=n$. Thus, if $p_i=\delta_{ik}$ for a certain $k$  
(maximum certainty) $\Rightarrow$ ${\bf q} \prec {\bf p}$ {\it for any} distribution ${\bf q}$, whereas 
for a uniform distribution $q_i=1/n$ $\forall$ $i$ (maximum uncertainty or mixedness) we have ${\bf q} \prec {\bf p}$ {\it for any} distribution ${\bf p}$. 
The concept of majorization also applies to probability distributions ${\bf p}=(p_1,\ldots,p_n)$ and ${\bf q}=(q_1,\ldots,q_m)$  with $m\neq n$, after padding with zeros  that with  the lesser number of elements. 

 If there exists a set of permutation matrices $\{\Pi_i\}$ with probabilities $r_i\geq 0$, $ \sum_i{r_i=1}$ such that ${\bf q}=\sum_i{r_i \Pi_i {\bf p}}$, i.e.\ if ${\bf q}$ is a convex combination of permutations $\Pi_i$ of $ {\bf p}$,  then ${\bf q} \prec {\bf p}$ and viceversa \cite{Bh.97}:
\begin{equation}
{\bf q}\prec {\bf p}\Longleftrightarrow {\bf q}=
\sum_i r_i \Pi_i{\bf p}\,,\label{perm}
\end{equation}
or, equivalently,
\begin{equation}
{\bf q}\prec {\bf p}\Longleftrightarrow {\bf q}=D{\bf p},
\end{equation}
where $D$ is a doubly stochastic matrix \cite{Bh.97}.
For this reason, if ${\bf q}\prec {\bf p}$, it is also said that the distribution ${\bf q}$ is ``more mixed'' than ${\bf p}$.
Of course, given two distributions ${\bf p}$, ${\bf q}$, it may occur that ${\bf q}\nprec {\bf p}$ and ${\bf p}\nprec {\bf q}$, that is, neither ${\bf p}$ or ${\bf q}$ majorizes the other, entailing that majorization constitutes a partial order relation between probability distributions.

It can also be shown \cite{Bh.97} that if  ${\bf q}\prec{\bf p}$ then  ${\bf q}$ has a larger entropy than  ${\bf p}$      \begin{equation}  
{\bf q}\prec {\bf p} \Rightarrow S({\bf q}) \geq S({\bf p}). 
\label{mayor1}
\end{equation}
where $S({\bf p})=-\sum_i p_i\log p_i$ is the Shannon entropy. The converse relation obviously does not hold in general, i.e., $S({\bf q}) \geq S({\bf p}) \nRightarrow 
 {\bf q}\prec{\bf p}$, entailing that   
 the concept of disorder provided by majorization is more stringent than that implied by the Shannon entropy.  

On the other hand, if $F$ is  be the set of smooth concave  functions $f:[0,1]\rightarrow \mathbb{R}$ that satisfy $f(0)=f(1)=0$,  it can be shown that \cite{RC.03}
\begin{equation}
 {\bf q}\prec  {\bf p}\Longleftrightarrow \sum_i f(q_i)\geq \sum_i f(p_i)\;\;\forall\;f\;\in\;F\,. 
\label{fconcava}
\end{equation} Recall that  concavity means 
$f[\alpha_1 p_1+\alpha_2 p_2]\geq \alpha_1 f(p_1)+\alpha_2f(p_2)$ 
$\forall\,\alpha_1,\alpha_2\in[0,1]$ with $\alpha_1+\alpha_2=1$, with equality valid iff $p_1=p_2$ or $\alpha_1\alpha_2=0$.
For smooth $f$, concavity is equivalent to decreasing $f'(p)$ for $p\in (0,1)$.

For $f\in F$, the function \cite{W.78,Cap5Bet.13,RC.03,RC.99}
\begin{equation}
S_f({\bf p})=\sum_i f(p_i)\label{Sf}\end{equation}
is a measure of the degree of ``mixedness'' associated with the probability distribution ${\bf p}$, and can be denoted  as  a {\it generalized trace form entropy}. It satisfies $S_f({\bf p})\geq 0$
with $S_f({\bf p})=0$ only if $p_i=\delta_{ij}$ for a certain $j=1,\ldots,n$ (maximum certainty), while  $S_f({\bf p})$ is  maximum for $p_i=1/n$ $\forall$ $i$ (maximum uncertainty).
Furthermore, equation (\ref{fconcava}) can be written as \cite{RC.03}
\begin{equation}
{\bf q}\prec {\bf p}\Longleftrightarrow S_f({\bf q})\geq S_f({\bf  p})\;\;\;\forall f\in\;{\textit F}. 
\label{Sconcava}
\end{equation}
Thus, 
if ${\bf q}\prec {\bf p}$
then $S_f({\bf q})\geq S_f({\bf p})$ {\it $ \forall$ $S_f$,  and conversely,  if $S_f({\bf q})\geq S_f({\bf p})$ $\forall$ $S_f$} then ${\bf q}\prec {\bf p}$.
That is, they are sufficient to ensure the majorization relation.
On the other hand, obviously $S_f({\bf q})>S_f({\bf p})$ for a given $f$ does not necessarily imply ${\bf q}\prec {\bf p}$. The majorization relation cannot be captured by a single entropy, and is therefore stricter than the ``disorder'' relation that emerges from a single type of entropy. In general, ${\bf q}\prec{\bf p}$ implies $S({\bf q})\geq S({\bf p})$ for any Schur- concave function $S$ of the probabilities 
\cite{Bh.97,MOA.11}.   

The extension of the concept of majorization to quantum density operators is straightforward. Given two density operators ($\rho\geq 0$, ${\rm Tr}\,\rho=1$), we say that $\rho'\prec\rho$ ($\rho'$ is {\it more mixed} than $\rho$) iff 
its eigenvalue spectrum $\bm\lambda(\rho')$, sorted in decreasing order, is majorized by that of $\rho$: 
\begin{subequations}
\begin{equation}    \rho'\prec\rho\;\Longleftrightarrow\;\bm\lambda(\rho')\prec\bm\lambda(\rho)\,.\label{prec01}
\end{equation}
Explicitly, 
\begin{equation}
\rho'\prec\rho\;\;\Longleftrightarrow \sum_{k=1}^l\lambda'_k\leq \sum_{k=1}^l\lambda_k\,,\;\;l=1,\ldots,n-1,\label{prec1}
\end{equation}
\label{prec11}
\end{subequations}
$\!\!$where $\lambda_k,\lambda_k'$ denote the  eigenvalues of $\rho$ and $\rho'$ respectively, sorted in {\it decreasing} order ($\lambda_{k+1}\leq \lambda_k$ $\forall\,k$) and $n$ is  the dimension of $\rho$ and $\rho'$ (${\rm Tr}\rho={\rm Tr}\rho'=1$ is assumed for $l=n$).

All previous entropic properties remain then valid for operators, replacing $S({\bf p})\longrightarrow S(\rho)$. Again, 
if the corresponding Hilbert space dimensions differ, the definition is applied by padding with zeros the spectrum of lowest dimension. 
Moreover, Eq.\ \eqref{perm} is generalized to \cite{Nielsen.01}
\begin{equation}    \rho'\prec\rho\;\Longleftrightarrow\;\rho'=\sum_i r_i U_i\rho U_i^\dag
\end{equation}
where $\{r_i\geq 0\}$ are again a set of  probabilities ($\sum_i r_i=1$) and $\{U_i\}$  a set of unitary operators. 
Eq.\ \eqref{Sconcava} implies \begin{equation}
\rho'\prec \rho\Longleftrightarrow S_f(\rho')\geq S_f(\rho)\;\;\;\forall f\in\;{\textit F}. 
\label{Sconc2}
\end{equation}

{\it Proof of Lemma 1}.
We  prove here Lemma 1 of sec.\ \ref{IID}, i.e.\ 
the sufficient condition for Eqs.\ \eqref{major}-\eqref{maj}, concerning the normalized system-time entanglement spectrum $\bm\lambda^{(N)}=\{\lambda_k^{(N)}\}$ of $|\Psi^{(N)}\rangle$ and 
$\bm\lambda^{(N+1)}=\{\frac{N}{N+1}\lambda_k^{(N)}\}\cup\{\frac{1}{N+1}\}$ of   $|\Psi^{(N+1)}\rangle$,  after the addition 
of a state $|\psi_n\rangle$ orthogonal to all previous states of the history at step $N$. 

After sorting the eigenvalues in decreasing order, the partial sums $S_k=\sum_{k'=1}^k \lambda_{k'}$  of $\bm\lambda^{(N)}$ and $\bm\lambda^{(N+1)}$ satisfy   $S^{(N+1)}_k=\frac{N}{N+1}S^{(N)}_k\leq S^{(N)}_k$ for $\lambda^{(N)}_k\geq \frac{1}{N}$, 
and $S^{(N+1)}_{k}=
\frac{N}{N+1}[S^{(N)}_{k}+\frac{1}{N}-\lambda^{(N)}_k]$ otherwise, again satisfying $S^{(N+1)}_{k}\leq S^{(N)}_k$  since 
$1-N\lambda_k^{(N)}\leq S^{(N)}_k$.
Hence $\bm\lambda^{(N+1)}\prec\bm\lambda^{(N)}$, which leads to Eqs.\ \eqref{major}-\eqref{maj}. \qed  
\\

{\it Example for regime II}. 
Let us consider  a sequence where one state $|\psi\rangle$ which is orthogonal to all other states of the history occurs $M<N$ times,  then leading to an eigenvalue $M/N$ of ${\bf O}_N$ which we assume is the largest.  If $\tilde{\bm\lambda}^{(N-M)}$ denotes the spectrum of  the overlap matrix of the remaining $N-M$ states, 
the sorted spectrum of ${\bf O}_N$ is 
 $\bm \lambda^{(N)}=(\frac{M}{N},\frac{\bm\tilde\lambda^{(N-M)}}{N})$, with $\tilde\lambda^{(N-M)}_{\rm max}\leq M$. Then  
further  addition of the same  $|\psi\rangle$  leads to a spectrum $\bm \lambda^{(N+1)}=(\frac{M+1}{N+1},\frac{\bm\tilde\lambda^{(N-M)}}{N+1})$ (discarding null eigenvalues).  It is easily verified that  $\bm\lambda^{(N)}\prec \bm\lambda^{(N+1)}$, since their  partial sums satisfy $S^{(N)}_{k+1}=\frac{M+\tilde S^{(N-M)}_k}{N}\leq S^{(N+1)}_{k+1}= \frac{M+1+\tilde S^{(N-M)}_k}{N+1}$, as $M+\tilde S_k^{(N-M)}\leq N$ for $k\leq N-M$, thus leading to  Eq.\ \eqref{minr}. 
If the last  state added does not coincide exactly with $|\psi\rangle$ but is sufficiently close, 
by continuity this regime will occur approximately (i.e., most entropies will decrease) since the last  eigenvalue added   will be very small.  
 
\section{Volatility}
\label{ApB}

Volatility, broadly understood as the degree of variation in asset returns, is a key variable in modern finance due to its impact on derivative pricing, portfolio construction and risk management. 
Its historical behavior offers valuable insights into the mechanisms governing market dynamics. Empirical studies have 
documented that volatility is not constant over time, but instead exhibits patterns such as clustering, mean reversion and asymmetric responses to market shocks. These facts, first formalized in models such as the Autoregressive Conditional Heteroskedasticity (ARCH) framework of \citet{engle1982arch} and its generalized form (GARCH) 
\citet{bollerslev1986generalized}, highlight the importance of analyzing past volatility as a means of understanding and forecasting future uncertainty. These models capture the empirical regularity of volatility clustering: periods of high volatility tend to be followed by high volatility, while those of low volatility tend to be followed by low volatility.

The study of past volatility is not solely of academic interest; it has immediate implications for market participants and policymakers, such as portfolio managers and derivative traders. Regulators and central banks may also rely on historical volatility analysis when designing stress-testing frameworks or assessing systemic risk, as elevated or persistent volatility can signal vulnerabilities in financial stability. In this context, the systematic examination of past volatility serves as both a diagnostic tool for identifying latent risks and a foundation for developing predictive models \cite{EPM.07} that inform decision-making across a wide spectrum of financial activities.

Despite decades of research, the analysis of past volatility remains highly relevant in today’s financial landscape. Structural shifts in market microstructure, the proliferation of algorithmic and high-frequency trading (HFT), and the emergence of new asset classes such as cryptocurrencies have altered the dynamics of price formation and volatility transmission. Moreover, periods of extreme market stress, such as the global financial crisis of 2008 or the COVID-19 pandemic, have highlighted the need for models capable of adapting to rapidly changing environments and capturing non-linear responses. 

{\it Volatility indicators}. 
The VIX index \cite{VIX.09} was introduced by the Chicago Board Options Exchange (CBOE) in 1993. It was originally designed to measure the implied volatility of at-the-money (ATM) SP-100 Index (OEX) option prices for 30 days \cite{whaley1993derivatives}. The VIX is now based on the S\&P500 index (or SPX),
the core index for U.S.\ equities, and estimates expected volatility by averaging the weighted quotes of SPX put and call options over a wide range of strike prices \cite{osterrieder2019vix}.
The VIX is presently the standard measure of volatility, and it even has ETFs that track its short-term futures, such as the ProShares VIX Short-Term Futures ETF (VIXY). Similarly,
the crypto volatility index (CVI),  also denoted as ``VIX for crypto'',  is an index elaborated by a decentralized platform  that measures the 30-day implied volatility of Bitcoin (BTC) and Ethereum (ETH) \cite{CVI}. 

%
\end{document}